\documentclass[review,12pt]{elsarticle}
\usepackage{lineno,hyperref}
\usepackage{graphicx}
\usepackage{geometry}
\usepackage{caption}
\usepackage{subcaption}
\usepackage[linesnumbered,ruled,lined]{algorithm2e}
\usepackage{float}
\usepackage{tikz}
\usepackage{epstopdf}
\usepackage{amssymb}
\usepackage{amsmath}
\usepackage{indentfirst}
\usepackage{subcaption}
\usepackage{footnote}
\usepackage{array}
\usepackage{cases}
\usepackage{multirow}
\usepackage{ upgreek }
\usepackage{makecell}
\begin{document}
%
%
\begin{frontmatter}
\title{The efficient implementation of transport velocity formulation}

\author[1]{Zhentong Wang}
\ead{zhentong.wang@tum.de}

\author[1]{Oskar J. Haidn}
\ead{oskar.haidn@tum.de}

\author[1]{Xiangyu Hu\corref{cor1}}
\ead{xiangyu.hu@tum.de}

\address[1]{TUM School of Engineering and Design, Technical University of Munich, Garching, 85748, Germany}

\cortext[cor1]{Corresponding author: Xiangyu Hu}

\begin{abstract}
The standard smoothed particle hydrodynamics (SPH) method suffers from tensile instability, 
resulting in particle clumping and void regions under negative pressure conditions. 
In this study, 
we extend the transport-velocity formulation of Adami et al. (2013) \cite{adami2013transport} in the weakly-compressible SPH (WCSPH) framework to address this long-standing issue. 
Rather than relying on background pressure, 
our modified and improved transport-velocity correction scales directly to the smoothing length, 
making it suitable for variable-resolution flows. 
Additionally, 
we introduce a limiter to the new formulation to prevent overcorrection, especially for flow with small velocities. 
These modifications enhance the general applicability of the transport velocity in fluid dynamics. 
Numerical tests involving low-velocity and variable-resolution cases demonstrate that the new formulation offers a general and accurate solution for multi-physics SPH simulations.
\end{abstract}

\begin{keyword}
Weakly-compressible SPH  \sep Tensile instability  \sep Smoothing length \sep Variable-resolution flows \sep Transport-velocity formulation
\end{keyword}
\end{frontmatter}
%
%
\section{Introduction}\label{Introduction}
Smoothed particle hydrodynamics (SPH), as a mesh-free computational technique initially proposed by Lucy \cite{lucy1977numerical}, Gingold and Monaghan \cite{gingold1977smoothed}, 
was developed for astrophysical simulations and has subsequently gained significant interest in various fields, including compressible \cite{monaghan2002sph,welton1998two} and incompressible fluid dynamics \cite{monaghan1994simulating}, 
solid mechanics \cite{libersky1993high}, 
and numerous scientific and industrial applications \cite{randles1996smoothed,longshaw2015automotive,wang2023eulerian,khayyer2022systematic}. 
The SPH method utilizes a kernel function within the compact support to approximate field quantities at arbitrarily distributed discretization elements (particles) through particle interactions and to dynamically update their positions based on local velocity.
Two typical strategies for solving incompressible fluid dynamics are employing the true incompressible SPH (ISPH) and weakly-compressible SPH (WCSPH) models. 
The former involves solving a pressure Poisson equation and enforcing the divergence condition for the velocity field. 
In contrast, 
the latter treats the fluid as weakly compressible with an artificial equation of state. 
Based on this, 
the WCSPH model has much higher computational efficiency compared with the ISPH model.
In the present study, 
we focus exclusively on the WCSPH framework.

When static pressure becomes negative, 
the SPH method experiences tensile instability in fluid simulations, 
leading to particle clumping or forming void regions.
This issue is particularly pronounced within the WCSPH framework, 
where minor density variations can cause significant pressure fluctuations.
Many techniques have been developed to address this problem.
Monaghan \cite{monaghan2005smoothed} introduced an artificial viscosity to prevent particles from clumping when in close proximity. 
However, this approach generates substantial dissipation, which strongly affects simulation outcomes.
Additionally, Schussler and Schmitt \cite{schuessler1981comments} proposed a new kernel function capable of generating a repulsive force under negative pressure. However, 
its function fails to meet the criterion for continuous first-order derivatives. 
Jonson and Beissel \cite{johnson1996normalized} employed a modified non-zero quadratic kernel function, 
although this kernel does not possess a continuous second-order derivative.
Furthermore, 
Colagrossi and Landrini \cite{colagrossi2003numerical} proposed utilizing the moving least squares (MLS) method to filter the density, thereby achieving a more stable pressure field at the cost of reduced computational efficiency.
Moreover, 
the particle shifting technique (PST), 
initially proposed by Monaghan \cite{monaghan1989problem} and further explored by Xu et al. \cite{xu2009accuracy, xu2018technique} and Lind et al. \cite{lind2012incompressible} within the ISPH framework, 
has been introduced in the WCSPH framework \cite{antuono2010free,sun2017deltaplus} and then applied in a variable-resolution scheme \cite{vacondio2013variable}.
Nevertheless, certain numerical issues arise due to its special free-surface treatment \cite{lyu2022further}.
To solve these numerical issues, 
an improved PST in the same framework has been introduced \cite{lyu2022further} but causing much more computational effort. 

One of the most popular approaches to address the particle clumping and void-region problems is introducing a transport velocity proposed by Adami et al. \cite{adami2013transport}.
By utilizing a globally uniform background pressure for regulation, 
the transport velocity enhances particle distribution with a kick-drift-kick time-integration scheme \cite{zhang2017weakly}, 
thereby improving zero-order consistency and reducing numerical error \cite{litvinov2015towards}.
A challenge to this transport-velocity formulation is its need for more straightforward applicability to problems involving free fluid or solid material surfaces.
Therefore, 
Zhang et al. \cite{zhang2017generalized} introduced a generalized transport-velocity formulation employing a variable background pressure to address the tensile instability issue for solid and fluid dynamics involving free surfaces.
Additionally, 
while dual-criteria time integration has been introduced in the WCSPH framework for enhancing computational efficiency, 
the transport velocity formulation relying on the background pressure \cite{zhang2020dual} with this time integration requires further clarification and discussion.
Except this limitation, 
the transport velocity method still encounters other several challenges. 
Firstly, 
since the correction of transport velocity is introduced to mitigate clumping and void-region issues, which directly reflect on particle distribution. 
Therefore, 
this correction is intuitively and straightforwardly relevant to adjust the particle's distance, scaled to the smoothing length rather than relying on the original formulation's background pressure \cite{adami2013transport}.
Secondly, the correction of transport velocity ought to remain  relatively minor compared to the updated displacement of particles in each time step. 
While significant accuracy improvements are evident in high-velocity scenarios \cite{adami2013transport,zhang2017generalized}, 
the original formulation may result in overcorrection in low-velocity simulations, which has yet to be discussed and solved \cite{adami2013transport,zhang2017generalized}. 
Finally, 
as this correction formulation is limited by the single-resolution scheme, 
further exploration is necessary to develop a formulation suitable for variable-resolution scenarios.

This paper presents the implementation of a modified transport-velocity formulation to address the tensile instability problem even involving the low-velocity flow.
This new formulation is derived from the original formulation relying on the background pressure framework proposed by Adami et al. \cite{adami2013transport} and directly  scaled to the smoothing length, 
which, therefore, has a clear advantage of being adapted for variable-resolution schemes with dual-criteria time integration. 
Additionally, 
a limiter included in the formulation is used to avoid overcorrection by the transport velocity in the low-velocity flow.
A series of numerical cases are conducted to demonstrate the proposed formulation's stablity and accuracy.
This paper is organized as follows: 
Sections \ref{WCSPH method} discusses the WCSPH framework and \ref{transport_velocity_correction} details the new proposed formulation of transport velocity correction. 
Besides, Section \ref{Numerical results} presents several numerical tests, 
and brief concluding remarks in Section \ref{Summary and conclusions}.
All computational codes employed in this study are publicly accessible via the SPHinXsys repository \cite{zhang2021sphinxsys}, available at https://www.sphinxsys.org.
%
%
\section{WCSPH method}\label{WCSPH method}
%
%
\subsection{Governing equations}\label{Eulerian SPH governing equatioins}
Within the Lagrangian framework, 
the principles of mass and momentum conservation for fluid dynamics are formulated as follows: 
\begin{equation}\label{conservation-equation}
\left\{\begin{array}{l}
\frac{d \rho}{d t}=-\rho \nabla \cdot \mathbf{v} \\
\rho \frac{d \mathbf{v}}{d t}=-\nabla p +\eta \nabla^{2}\mathbf{v}+ \rho\mathbf{g}
\end{array},\right.
\end{equation}
where $\mathbf{v}$ denotes the velocity field, and $\rho$, $p$, $\eta$, and $\mathbf{g}$ represent the density, pressure, dynamic viscosity, and gravity force, respectively. The material derivative is defined as $\frac{d}{d t}=\frac{\partial}{\partial t} + \mathbf{v} \cdot \nabla$. 
For incompressible flows, 
based on the weakly-compressible assumption, 
an artificial equation of state (EOS) is introduced to close the system given by 
\begin{equation}
p=c^2(\rho-\rho_0). 
\end{equation}
Here, $\rho_0$ is the reference density, and the artificial sound speed $c=10U_{max}$, where $U_{max}$ is the maximum velocity of the flow, ensures that density variation is kept below 1\%.
%
%
\subsection{SPH discretization}\label{Eulerian SPH discretization}
Following Refs. \cite{zhang2017weakly,zhang2020dual}, 
the Riemann-based discretization of Eq. \eqref{conservation-equation} can be written as 
\begin{equation}\label{eqs:conservation-discretize}
\left\{\begin{array}{l}
\frac{d \rho}{d t}=2\rho_{i} \sum_{j} V_{j} (U^{*}_{ij}-\mathbf{v}_{i}\cdot \mathbf{e}_{ij}) \frac{\partial W_{ij}}{\partial r_{ij}} \\
\frac{d \mathbf{v}_{i}}{d t}=-2\sum_{j} m_{j} \frac{P^{*}_{ij}}{\rho_{i}\rho_{j}}\nabla W_{ij}
\end{array},\right.
\end{equation}
where $m$ and $V$ are the particle mass and volume, respectively, and 
the kernel gradient $\nabla W_{ij}=\frac{\partial W_{ij}}{\partial r_{ij}}\mathbf{e}_{ij}$ with $\mathbf{e}_{ij}=-\mathbf{r}_{ij}/r_{ij}$ where $\mathbf{r}_{ij}$ represents the displacement.
The terms $()^{*}_{ij}$ are determined by solving the Riemann problem.
For incompressible flows, 
the linearized Riemann solver \cite{toro2013riemann} is utilized, expressed as: 
\begin{equation}\label{linearised Riemann solver}
\left\{\begin{array}{l}
U^{*}_{ij}=\frac{U_{l}+U_{r}}{2}+\frac{1}{2} \frac{\left(p_{l}-p_{r}\right)}{\bar{\rho} \bar{c}} \\
P^{*}_{ij}=\frac{p_{l}+p_{r}}{2}+\frac{1}{2} \bar{\rho}  \bar{c} \left(U_{l}-U_{r}\right)
\end{array},\right.
\end{equation}
where $\bar{\rho}$ and $\bar{c}$ denote the particle averages, 
,$U_{l/r}=\mathbf{v}_{i/j}\cdot \mathbf{e}_{ij}$, and $p_{l/r}=p_{i/j}$.
To decrease the numerical dissipation,
a dissipation limiter is incorporated into the linearized Riemann solver as:  
\begin{equation}\label{acoustic Riemann solver}
\left\{\begin{array}{l}
U^{*}_{ij}=\frac{U_{l}+U_{r}}{2}+\frac{1}{2} \frac{\left(p_{L}-p_{R}\right)}{\bar{\rho} \bar{c}} \\
P^{*}_{ij}=\frac{p_{l}+p_{r}}{2}+\frac{1}{2} \beta \bar{\rho} \bar{c} \left(U_{l}-U_{r}\right)
\end{array}.\right.
\end{equation}
The dissipation limiter $\beta$ is defined as follows:
\begin{equation}\label{acoustic limiter}
\beta=\min \left(\upeta \max (\frac{U_{l}-U_{r}}{\bar{c}}, 0), 1\right),
\end{equation}
with $\upeta=3$.

For the viscous flows,
the physical shear term in Eq. \eqref{conservation-equation} is discretized \cite{hu2006multi} as 
\begin{equation}
(\frac{d \mathbf{v}_{i}}{d t})^{\nu}=2\sum_{j} m_{j} \frac{\eta}{\rho_{i}\rho_{j}}\frac{\mathbf{v}_{ij}}{r_{ij}} \frac{\partial W_{ij}}{\partial r_{ij}}.
\end{equation}
%
%
%
\subsection{Evolution of density}\label{Evolution of density}
In incompressible flows,
the density evolution, i.e. mass equation in Eq. \eqref{conservation-equation}, can lead to significant density errors, especially at high Reynolds numbers \cite{colagrossi2003numerical}. 
To address this issue, 
the density field is reinitialized \cite{zhang2020dual,zhang2021sphinxsys} using the following equation: 
\begin{equation}
	\rho_i=\rho^{0}\frac{\sum W_{ij}}{\sum W^{0}_{ij}}.
\end{equation}
where $\rho^{*}$ denotes the density prior to reinitialization, and the superscript $0$ indicates the initial reference value.
%
%
\subsection{Dual-criteria time stepping}
To enhance computational efficiency, we implement the dual-criteria time-stepping approach for integrating fluid equations. 
Following Refs. \cite{zhang2020dual,zhang2021sphinxsys}, 
there are two time-step size criteria: the advection criterion, denoted as $\Delta t_{ad}$, 
is defined by
\begin{equation}\label{advection_time}
\Delta t_{ad}=\text{CFL}_{ad}\frac{h}{|\mathbf{v}|_{ad}},
\end{equation}
with particle advection speed $|\mathbf{v}|_{ad}$ as 
\begin{equation}\label{max_velocity}
|\mathbf{v}|_{ad}=\text{max}(|\mathbf{v}|_{max},\text{max}(\frac{\nu}{h},U_{ref})),
\end{equation}
and the acoustic criterion, denoted as $\Delta t_{ac}$, 
is given by
\begin{equation}
\Delta t_{ac}=\text{CFL}_{ac}\frac{h}{|\mathbf{v}|_{max}+c}.
\end{equation}
with $\text{CFL}_{ad}=0.25$ and $\text{CFL}_{ac}=0.6$. 
Here, 
$|\mathbf{v}|_{max}$, $\nu$, $h$ and $U_{ref}$ denote the maximum particle fluid speed, the kinematic viscosity, smoothing length and reference velocity, respectively.
It is important to note that in variable-resolution flows, 
as the different initial particle spacing $dp$ across different regions,
the particle volumes $V=(dp)^{D}$ differ with $D$ denoting the dimensions, and the smoothing lengths, $h=\alpha dp$ with $\alpha$ a constant, are also different.
Note that, in Eq. \eqref{max_velocity}, 
the smoothing length involving variable resolutions is applied as the minimum value, 
i.e., $h=h_{min}=\alpha dp_{min}$.
Among the dual-criteria time-stepping approach, the advection criterion updates the topology relationship, while the acoustic criterion determines the update of fluid pressure and density.
%
%
\section{Transport velocity correction}\label{transport_velocity_correction}
To prevent particle clumping and void regions in the SPH method, 
the transport velocity \cite{zhang2017generalized,adami2013transport} is introduced as follows:
\begin{equation}\label{transport_equation}
\frac{d \widetilde{\mathbf{v}}_{i}}{d t}=\frac{d \mathbf{v}_{i}}{d t}-2\sum_{j} m_{j} \frac{p^{0}}{\rho_{i}\rho_{j}}\nabla W_{ij}.
\end{equation}
Here, 
the background pressure $p^{0}=\rho_{0}c^{2}$ depends on the sound speed in the kick-drift-kick scheme \cite{monaghan1992smoothed} for time integration and determines the degree of transport velocity correction. 
In the dual-criteria time-stepping scheme \cite{zhang2020dual}, 
the transport velocity correction is applied at each advection time step, 
where the background pressure is redefined as 
\begin{equation}\label{new_pressure}
p^{0}=\alpha \rho_{0}|\mathbf{v}|_{ad}^{2}, 
\end{equation}
with $\alpha$ set to $7.0$ \cite{zhang2023lagrangian}. 
It is important to note that, 
unlike the background pressure in the kick-drift-kick scheme, 
which depends on the sound speed, 
the background pressure here is dependent on the particle's advection velocity in this time-stepping scheme.

\begin{figure}
	\centering
	\begin{subfigure}[b]{0.49\textwidth}
		\centering
		\includegraphics[trim = 0cm 0cm 0cm 0cm, clip, width=1.0\textwidth]{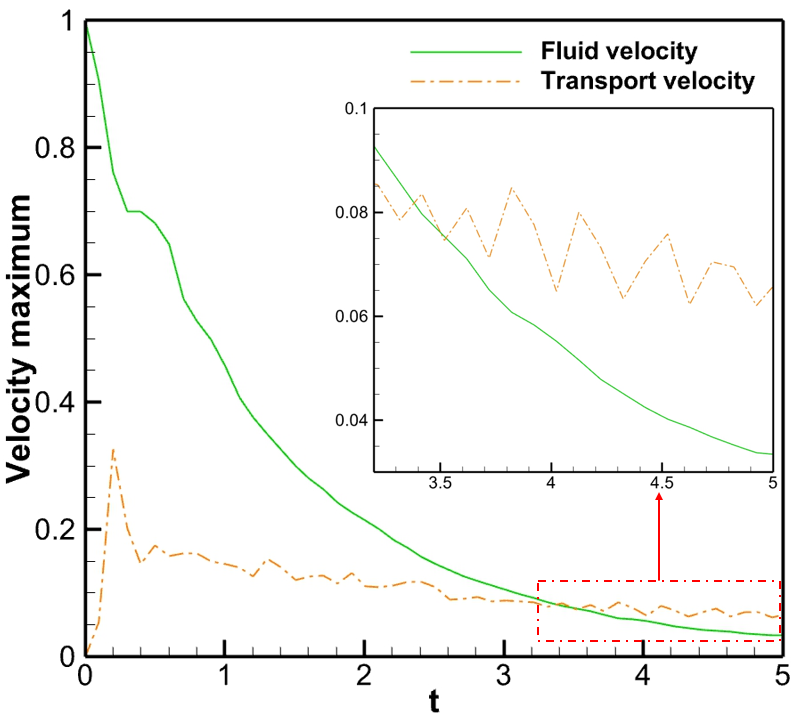}
	\end{subfigure}
	\begin{subfigure}[b]{0.49\textwidth}
		\centering
		\includegraphics[trim = 0cm 0cm 0cm 0cm, clip, width=1.0\textwidth]{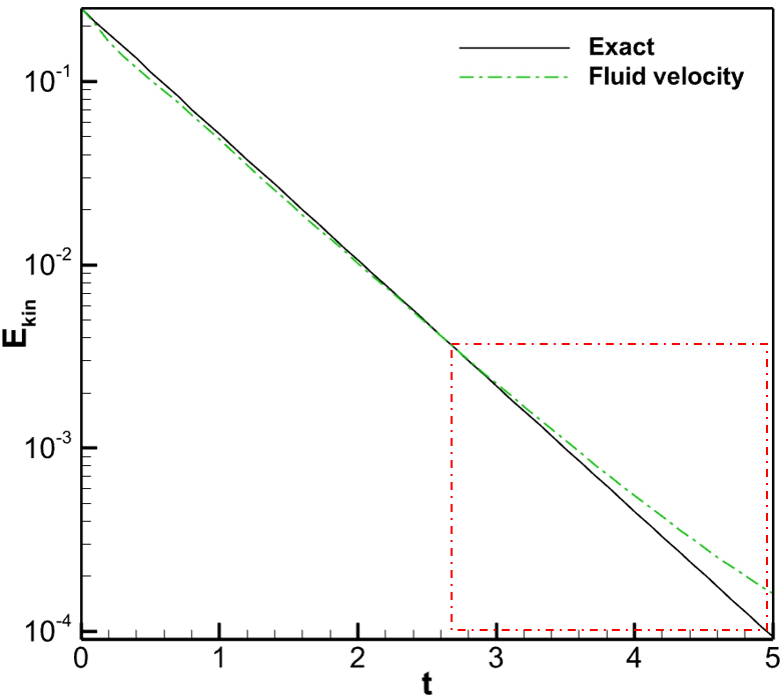}
	\end{subfigure}
	\caption{The maximum velocity (left panel) and energy decay (right panel) in the Taylor-Green vortex flow with the spatial resolution $dp=1/200$ and Reynolds number $Re=100$.}
	\label{Taylor-Green flow analysis}
\end{figure}

Upon inserting Eq. \eqref{new_pressure} into Eq. \eqref{transport_equation}, 
the displacement correction of the transport velocity at each advection timestep is formulated as 
\begin{equation}\label{new_transport_equation}
\Delta \widetilde{\mathbf{r}}=\frac{1}{2}\mathbf{a}_{t}(\Delta t_{ad})^{2}=\frac{1}{2}(-2 \sum_{j} m_{j} \frac{\alpha \rho_{0}|\mathbf{v}|_{ad}^{2}}{\rho_{i}\rho_{j}}\nabla W_{ij})(\text{CFL}_{ad}\frac{h}{|\mathbf{v}|_{ad}})^{2} =\xi h^{2}\sum_{j}\nabla W_{ij}V_{j},
\end{equation}
with the coefficient $\xi=-0.2$ accordingly.
Unlike the formulation in Eq. \eqref{transport_equation}, 
the new formulation in Eq. \eqref{new_transport_equation} is scaled and only dependent on the smoothing length and the particle distribution, i.e., zero-order consistency.
Despite using different scales, 
Eq. \eqref{new_transport_equation} remains inherently and algorithmically equivalent to Eq. \eqref{transport_equation}, 
ensuring that both formulations consistently achieve the high robustness and accuracy \cite{zhang2020dual,zhang2023lagrangian}. 
The reason for improved accuracy is that the correction of displacement, 
as relatively minor compared to the particle's updated displacement, 
is beneficial for achieving enhanced zero-order consistency.
However, 
an overlooked issue is that in low-velocity flows, 
the correction of Eqs. \eqref{transport_equation} and \eqref{new_transport_equation} can lead to an overcorrection, 
i.e., the transport velocity exceeds the fluid velocity.
To clearly illustrate this phenomenon,
we utilize the transport velocity formulation of Eqs. \eqref{transport_equation} and \eqref{new_transport_equation} in the two-dimensional Taylor-Green case with the unit square size discretized spatially as $1/200$ under the Reynolds number $Re=100$.
Figure \ref{Taylor-Green flow analysis} presents the maximum fluid and transport velocity magnitude and the kinetic energy decay.
The left panel shows that the maximum transport velocity exceeds the fluid velocity within the red dashed box, indicating overcorrection. 
This overcorrection leads to a smaller-than-expected energy decay as the unphysical result, as depicted in the right panel.

To address the overcorrection issue, 
it is essential to derive and implement an effective limiter.
The crucial point is that the magnitude of correction from the transport velocity depends on the degree of uniformity of the particle distribution. 
In a disordered particle distribution, 
a relatively large correction, 
not exceeding the original correction in Eq. \eqref{new_transport_equation}, 
is beneficial for achieving zero-order consistency. 
Conversely, 
in a well-ordered particle distribution, 
only a minor or no correction is employed to avoid overcorrection.
Therefore, 
we propose the following limiter:
\begin{equation}\label{limiter}
	\beta= \begin{cases}\min \left(C h^{2}|\sum_{j}\nabla W_{ij}V_{j}|^{2}, 1\right), &  h^{2}|\sum_{j}\nabla W_{ij}V_{j}|^{2} > 5\times10^{-4} \\ 0, & \text { otherwise }\end{cases}.
\end{equation}
with $C=10^{3}$ according to the numerical cases. 
Note that, 
for the flows involving the variable resolutions, 
the smoothing length $h$ in Eqs. \eqref{new_transport_equation} and \eqref{limiter} is $h_{min}$.
Finally, 
the improved transport velocity correction is written as 
\begin{equation}\label{new_transport_equation_limiter}
\Delta \widetilde{\mathbf{r}}=-0.2\beta h_{min}^{2}\sum_{j}\nabla W_{ij}V_{j}.
\end{equation}
%
%
%
\section{Numerical results}\label{Numerical results}
In this section, 
we examine a series of numerical cases, 
including Taylor-Green vortex flow, $2D$ and $3D$ lid-driven cavity problems, fluid-structure interaction (FSI), multi-resolution flow around a cylinder and $3D$ FDA nozzle, 
to assess the performance of the proposed formulation. 
For clarity, we employ the Wendland kernel \cite{wendland1995piecewise} with a smoothing length of $h=1.3dp$. 
%
%
\subsection{Taylor–Green vortex}
In this segment, 
we analyze the two-dimensional viscous Taylor-Green vortex flow to verify the effectiveness of the proposed formulation in avoiding overcorrection.
As detailed in Ref. \cite{taylor1937mechanism}, the initial velocity within a unit domain subject to periodic boundary conditions along both the $x-$ and $y-$ axes is specified as follows:
\begin{equation}\label{taylor green initial condition}
	\left\{\begin{array}{l}
	u(x,y,t)=-\exp^{bt}cos(2\pi x)sin(2\pi y) \\
	v(x,y,t)=\exp^{bt}sin(2\pi x)cos(2\pi y) 
	\end{array}.\right.
\end{equation}
Here, the decay rate is given by $b=-8\pi ^2 /Re$, 
with the Reynolds number $Re=100$. 
The total kinetic energy decay rate is $-16\pi ^2 /Re$, 
and the final time is $t=5$.

\begin{figure}
	\centering
	\includegraphics[width=0.95\textwidth]{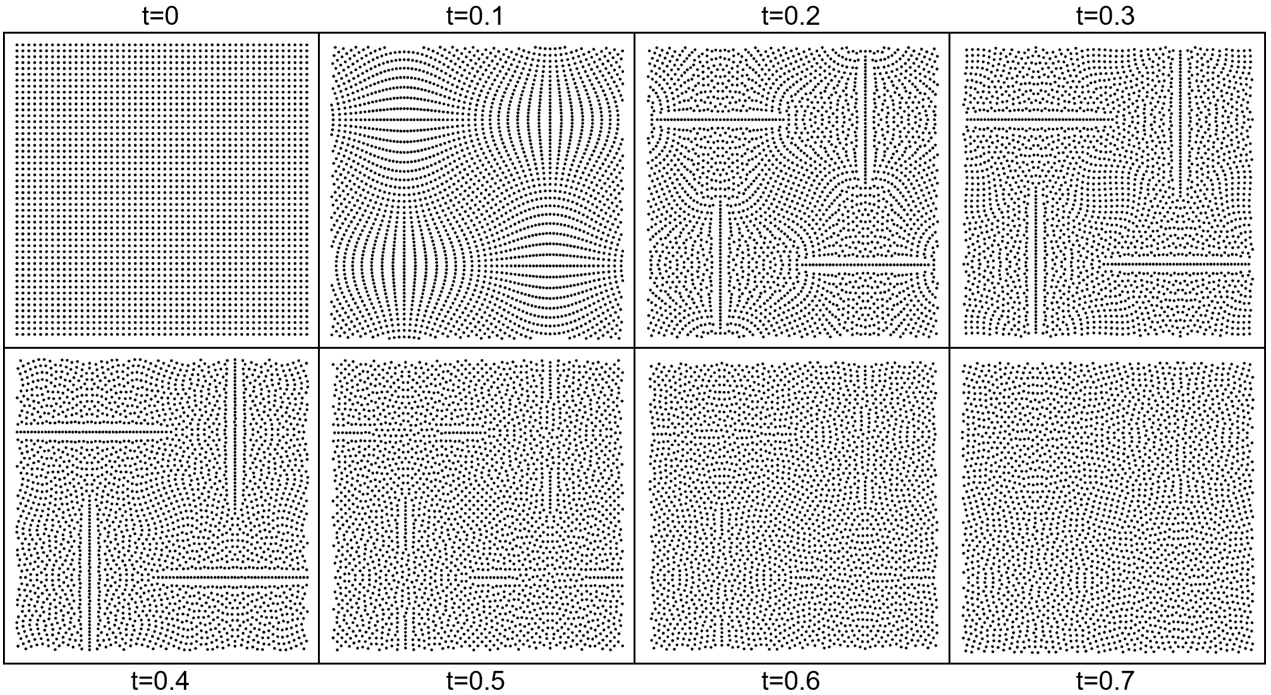}
	\caption{Taylor–Green vortex ($Re = 100$): Particle snapshots with a resolution of $50 \times 50$ particles at different time instants.}
	\label{Taylor-Green_particle_distribution}
\end{figure}
\begin{figure}
	\centering
	\begin{subfigure}[b]{0.49\textwidth}
		\centering
		\includegraphics[trim = 0cm 0cm 0cm 0cm, clip, width=1.0\textwidth]{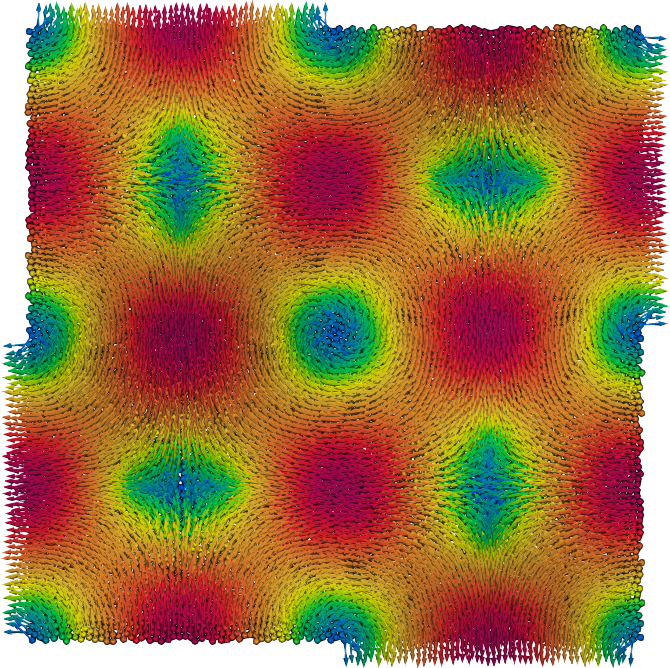}
	\end{subfigure}
	\begin{subfigure}[b]{0.49\textwidth}
		\centering
		\includegraphics[trim = 0cm 0cm 0cm 0cm, clip, width=1.0\textwidth]{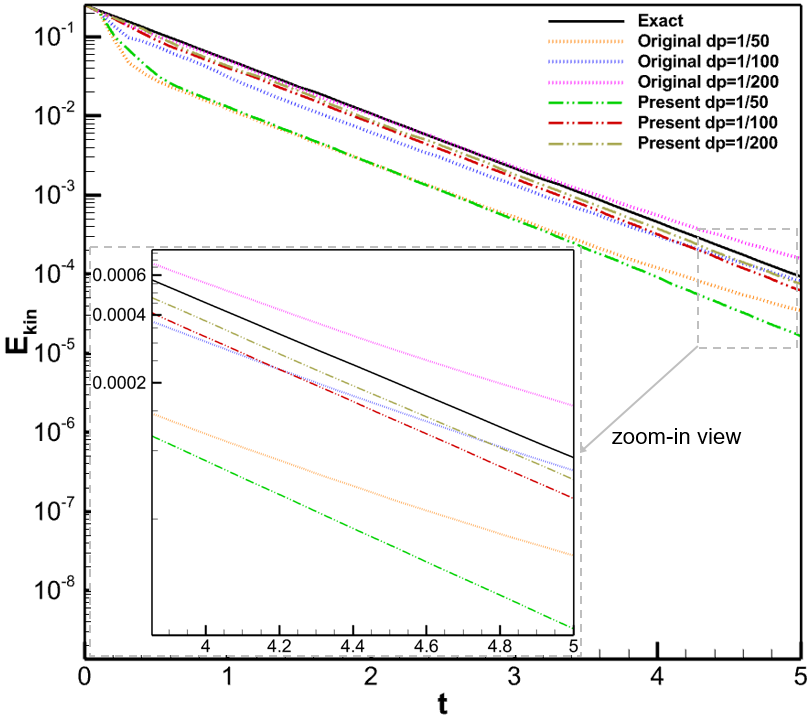}
	\end{subfigure}
	\caption{Taylor–Green vortex ($Re = 100$): 
 Instantaneous velocity field and velocity vectors with a resolution of $dp=1/100$ (left panel) and decay of kinetic energy (right panel).}
	\label{Taylor-Green_velocity_distribution}
\end{figure}
Figure \ref{Taylor-Green_particle_distribution} illustrates particle distributions at various time intervals with a resolution of $50 \times 50$ particles. 
It is evident that the particles, 
originating from an initial regular lattice configuration, 
are produced with a homogeneous particle distribution devoid of clustering.
Figure \ref{Taylor-Green_velocity_distribution} depicts the velocity magnitude, velocity vectors, and kinetic energy decay.  
The left panel employs the velocity vectors to facilitate a clear depiction of the flow structure.  
In the right panel, 
the decay of kinetic energy with the original formulation exhibits overcorrection at three different resolutions.  
In contrast, the proposed formulation is observed to align with the decay of kinetic energy with the analytical solution, 
validating that it effectively avoids the overcorrection issue.
%
%
\subsection{Lid-driven cavity flow}
Although the primary goal of the proposed method is not to enhance numerical accuracy, 
assessing its numerical accuracy remains essential. 
Consequently, 
in this section, 
we examine two-dimensional lid-driven cavity flow to evaluate the accuracy of the proposed formulation.
The computational domain is a square with a length of $L=1$, 
where the top wall moves at a constant speed $U_{max}=1$, 
and the Reynolds number $Re=1000$.

\begin{figure}
	\centering
	\begin{subfigure}[b]{0.42\textwidth}
		\centering
		\includegraphics[trim = 0cm 0cm 0cm 0cm, clip, width=1.0\textwidth]{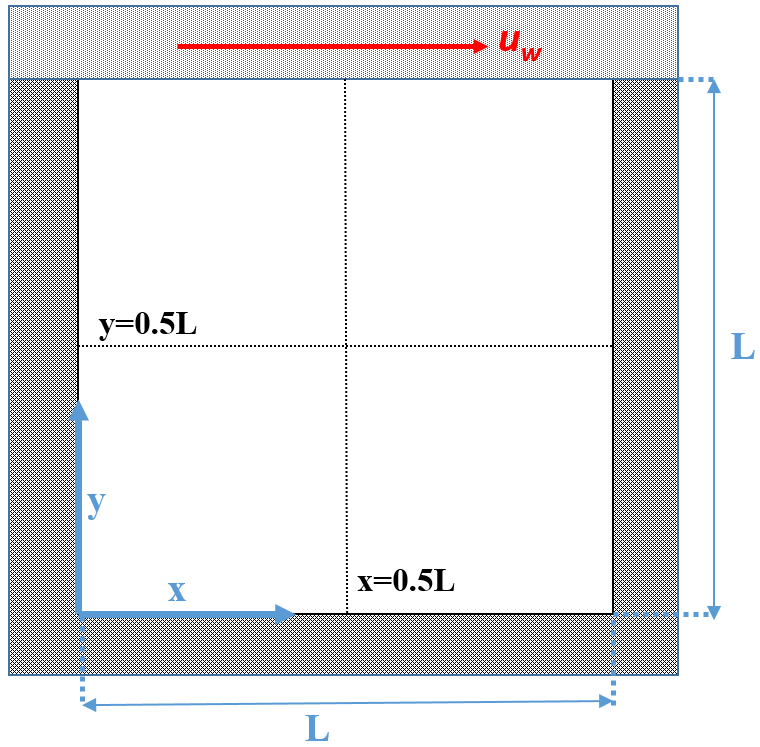}
	\end{subfigure}
	\centering
	\begin{subfigure}[b]{0.55\textwidth}
		\centering
		\includegraphics[trim = 0cm 0cm 0cm 0cm, clip, width=1.0\textwidth]{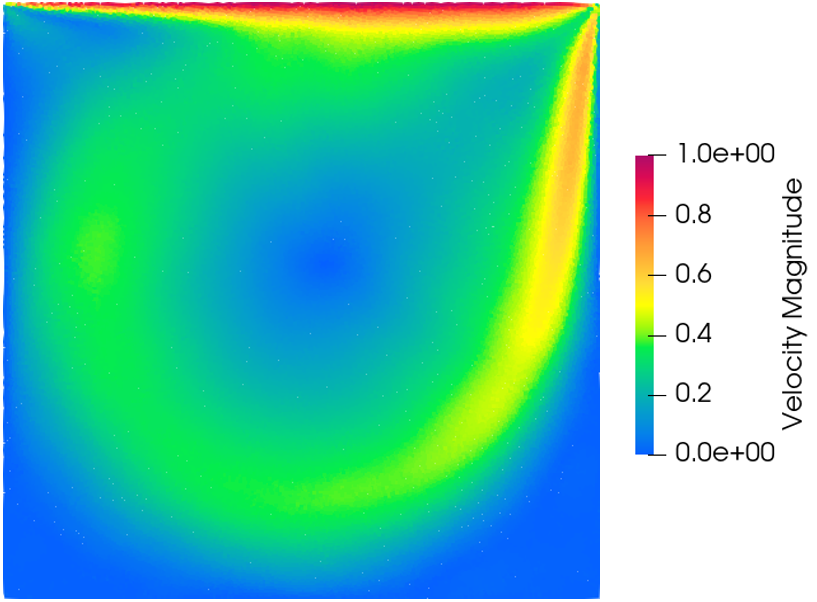}
	\end{subfigure}
	\caption{Lid-driven square cavity flow ($Re=1000$): The geometry and boundary conditions setup and the velocity contour ranging from $0$ to $1$.}
	\label{lid-driven-setup-and_contour}
\end{figure}
\begin{figure}
	\centering
	\begin{subfigure}[b]{0.49\textwidth}
		\centering
		\includegraphics[trim = 0cm 0cm 0cm 0cm, clip, width=1.0\textwidth]{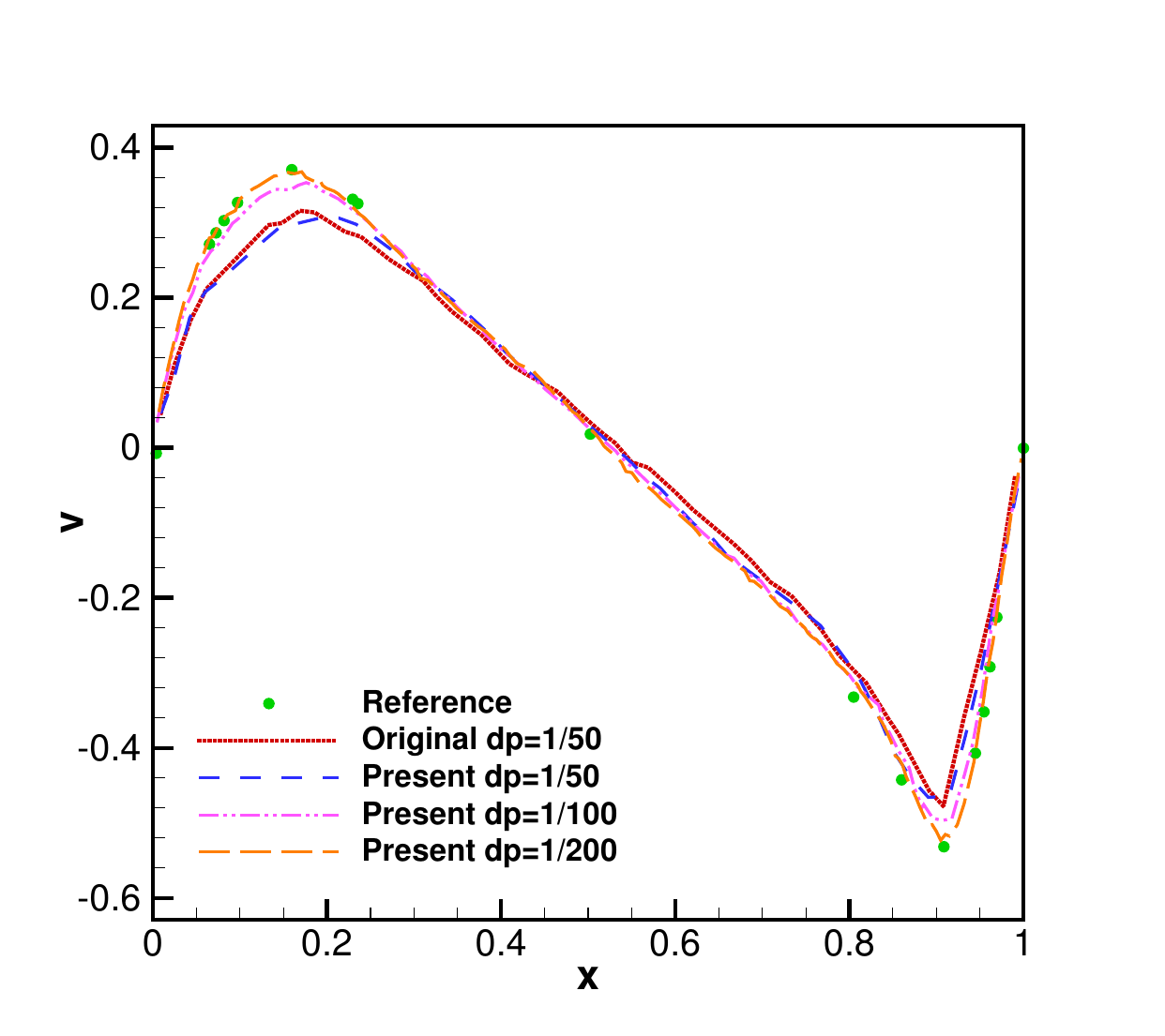}
	\end{subfigure}
	\centering
	\begin{subfigure}[b]{0.49\textwidth}
		\centering
		\includegraphics[trim = 0cm 0cm 0cm 0cm, clip, width=1.0\textwidth]{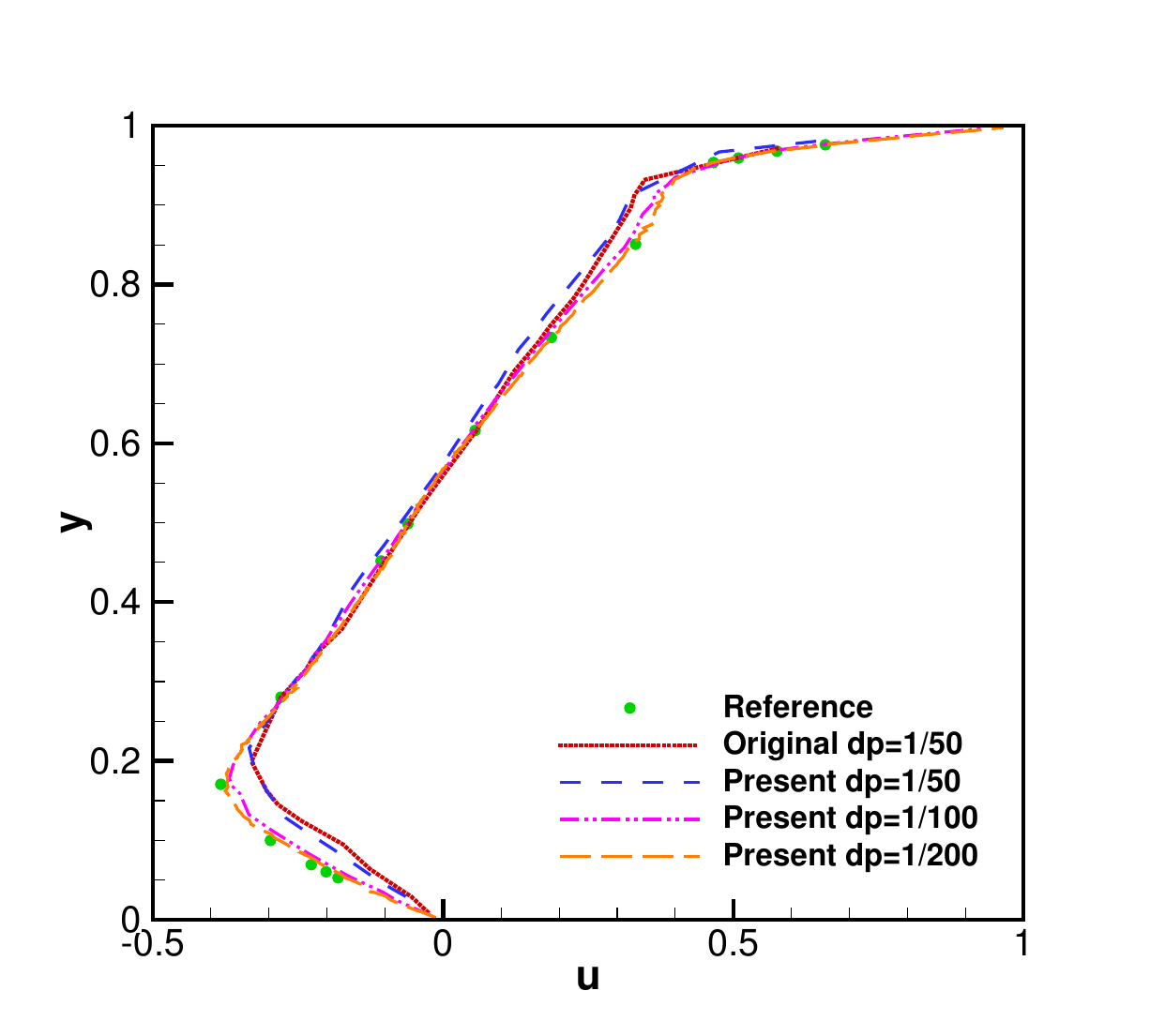}
	\end{subfigure}
	\caption{Lid-driven square cavity flow ($Re=1000$): The horizontal velocity component $u$ along $x = 0.5L$ (left panel) and 
		the vertical velocity component $v$ along $y = 0.5 L$ (right panel) with the spatial resolutions as $dp=1/50$, $1/100$ and $1/200$, and the comparisons with the reference obtained by Ghia et al. \cite{ghia1982high}.}
	\label{lid-driven-cavity-velocity_profile}
\end{figure}
Figure \ref{lid-driven-cavity-velocity_profile} illustrates the horizontal velocity component $u$ at $x = 0.5L$ and the vertical velocity component $v$ at $y = 0.5L$ using spatial resolutions of $dp=1/50$, $1/100$ and $1/200$, 
along with a comparison to the reference data from Ghia et al. \cite{ghia1982high}. 
The comparison demonstrates that the results converge rapidly, 
showing an approximate second-order convergence as the resolution increases. 
Additionally, 
the findings confirm that the proposed formulation maintains numerical accuracy consistent with the original formulation.
%
%
\subsection{Flow-induced vibration of an elastic beam behind a cylinder}
\begin{figure}
	\centering
	\includegraphics[width=0.6\textwidth]{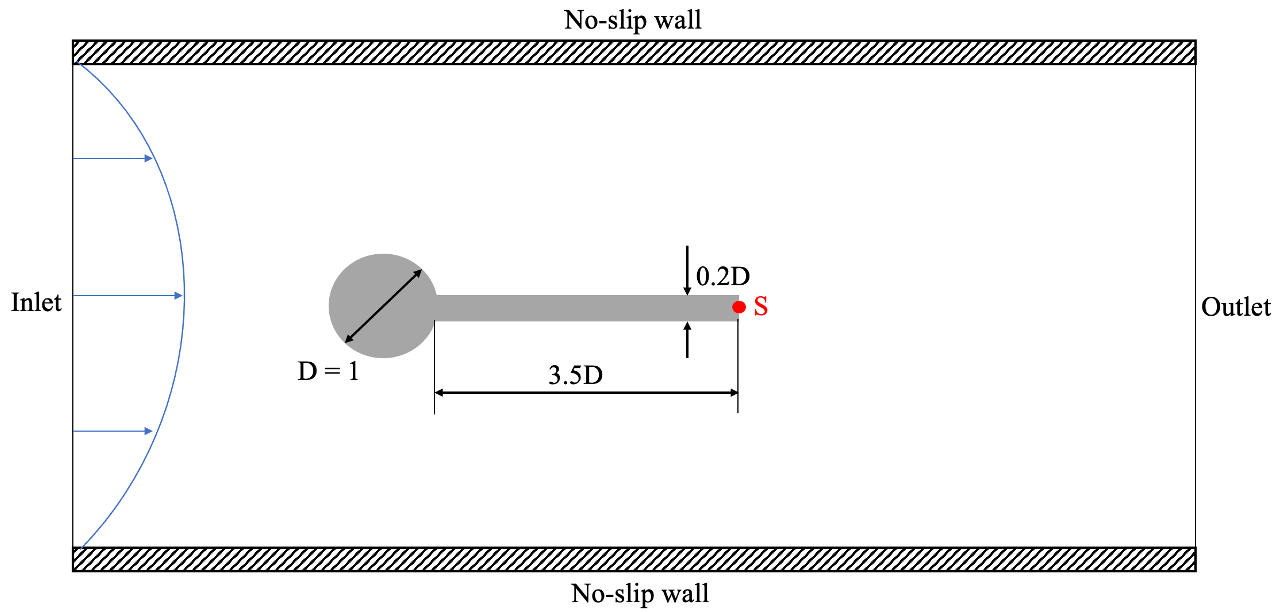}
	\caption{Sketch of the two-dimensional flow-induced oscillation of an elastic beam attached to a cylinder with a trajectory sensor S located at the free-end of the beam.}
	\label{FSI_geometry_and_boundary}
\end{figure}

This section tests the two-dimensional flow-induced vibrations of a flexible beam attached to a rigid cylinder to assess the effectiveness of the proposed formulation in the FSI problem. 
The geometric parameters and fundamental setup \cite{turek2006proposal,han2018sph} of the problem are depicted in Figure \ref{FSI_geometry_and_boundary}.
The rigid cylinder is positioned at coordinates $(2D, 2D)$ with $D=1$ measured from the lower-left corner of the computational domain. 
No-slip boundary conditions are enforced on the top and bottom,
while the left and right boundaries implement inflow and outflow conditions, respectively. 
The inflow condition follows a velocity profile \cite{turek2006proposal,han2018sph} given by $U(y)=1.5\overline{U}(t,y)(H-y)y/H^{2}$, where $\overline{U}(t,y)=0.5U_{0}(1.0-\cos{0.5\pi t})$ if $t\leq t_{ref}$ otherwise $\overline{U}(t,y)=U_{0}$, with $U_{0}=1.0$ and $t_{ref}=2.0$.
The density ratio of the structure to the fluid is $\rho_f /\rho_s=1/10$. 
The dynamic viscosity is calculated by $\mu = \rho_f U_{0}D/Re$, 
where the Reynolds number $Re = 100$. 
Dimensionless Young’s modulus $E^{\ast} = E/(\rho_f U^2_{0}) = 1.4 \times 10^3$, and Poisson ratio $\nu^s = 0.4$.

\begin{figure}
	\centering
	\includegraphics[width=1.0\textwidth]{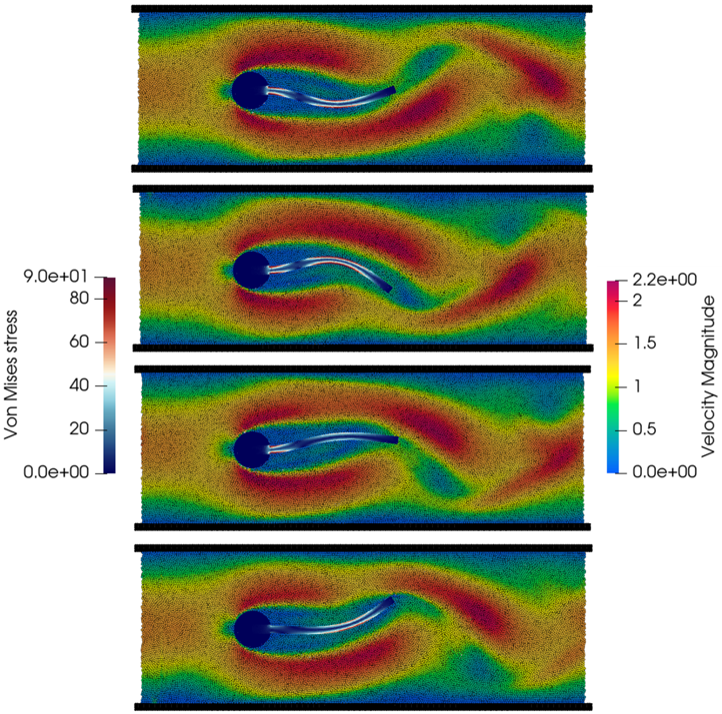}
	\caption{The fluid velocity magnitude contour at different instances, with the corresponding beam deformation colored by Von Mises stress.}
	\label{FSI_different_instants}
\end{figure}
\begin{table}[htbp]
\centering
\caption{Flow-induced vibration of an elastic beam attached to a cylinder: Comparison of oscillation amplitude in y-direction and frequency.}
\label{table:oscillation_comparison}
\begin{tabular}{ccc}
\hline
\textbf{Reference} & \textbf{Amplitude in y direction/(D)} & \textbf{Frequency} \\
\hline
Turek and Hron \cite{turek2006proposal} & 0.830 & 0.190 \\
Zhang et al. \cite{zhang2021multi} & 0.855 & 0.189 \\
Bhardwaj and Mittal \cite{bhardwaj2012benchmarking} & 0.920 & 0.190 \\
Tian et al. \cite{tian2014fluid} & 0.784 & 0.190 \\
Present & 0.790 & 0.186 \\
\hline
\label{FSI_table}
\end{tabular}
\end{table}
\begin{figure}
	\centering
	\begin{subfigure}[b]{0.9\textwidth}
		\centering
		\includegraphics[trim = 0cm 0cm 0cm 0cm, clip, width=1.0\textwidth]
        {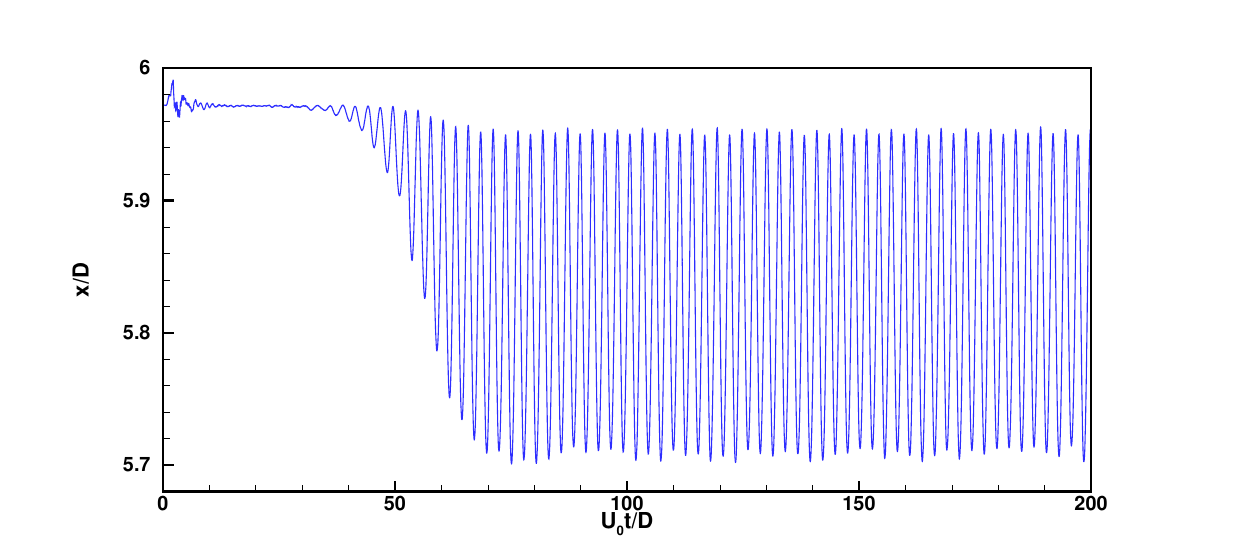}
	\end{subfigure}
	\begin{subfigure}[b]{0.9\textwidth}
		\centering
		\includegraphics[trim = 0cm 0cm 0cm 0cm, clip, width=1.0\textwidth]{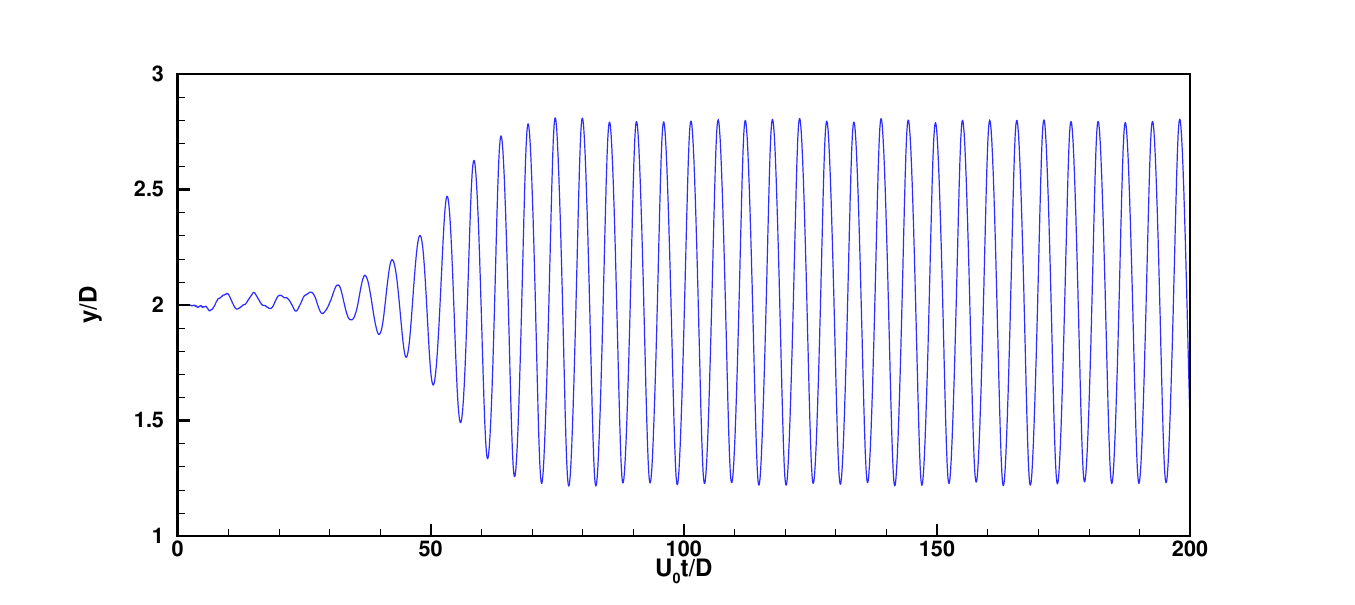}
	\end{subfigure}
	\centering
	\begin{subfigure}[b]{0.9\textwidth}
		\centering
		\includegraphics[trim = 0cm 0cm 0cm 0cm, clip, width=1.0\textwidth]{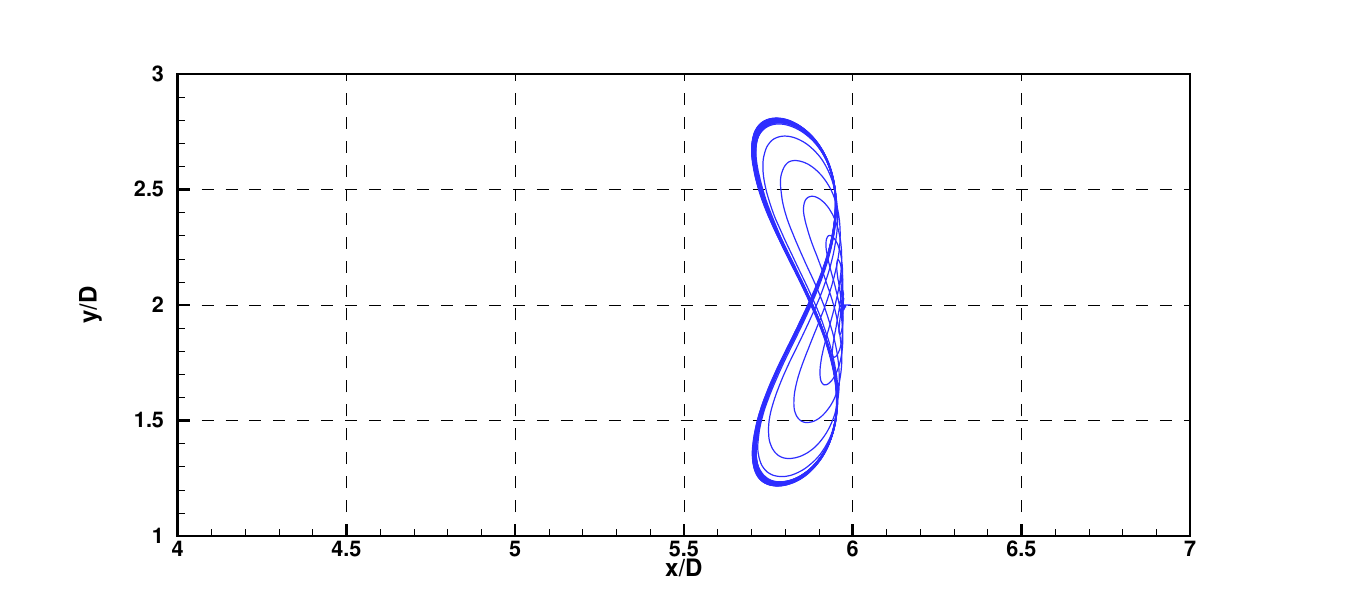}
	\end{subfigure}	
	\caption{Flow-induced vibration of an elastic beam attached to a cylinder: $x-$direction displacement (top panel), $y-$direction displacement (middle panel) and the trajectory (bottom panel) of point S.}
	\label{FSI_displacement}
\end{figure}

Figure \ref{FSI_different_instants} illustrates the flow vorticity field and beam deformation at various time points during a typical periodic movement after self-sustained oscillation is established. 
The results strongly agree with previous computational studies \cite{turek2006proposal,zhang2021multi}.
Figure \ref{FSI_displacement} illustrates the horizontal and vertical displacements of point S, 
located at the beam's fixed end, 
as shown in Figure \ref{FSI_geometry_and_boundary}. 
After the dimensionless time surpasses $U_{0}t/D \textgreater 70$, 
the beam reaches a state of periodic oscillation with a consistent amplitude and frequency. 
The movement path of point S forms a Lissajous curve, 
demonstrating a $2:1$ frequency ratio between the horizontal and vertical displacements. 
This behaviour is consistent with findings from Refs. \cite{turek2006proposal,zhang2021multi}.
Table \ref{FSI_table} lists the dimensionless amplitude of the oscillation in $ y-$ direction and the frequency, 
indicating that the proposed formulation produces convincing results.
%
%
\subsection{Multi-resolution flow around cylinder}
In this section, 
we examine the flow around a cylinder to demonstrate the flexibility of the proposed formulation within a variable-resolution framework. 
To quantitatively assess the numerical results, the drag and lift coefficients are defined as follows:
\begin{equation}\label{eq:wavespeed}
C_{D}=\frac{2F_{D}}{\rho_{\infty}u_{\infty}^2 A},    C_{L}=\frac{2F_{L}}{\rho_{\infty}u_{\infty}^2 A},
\end{equation}
where $F_{D}$ and $F_{L}$ represent the drag and lift forces exerted on the cylinder, respectively. 
Free-stream boundary conditions \cite{zhang2023lagrangian} are applied to all boundaries, with the field velocity $u_{\infty}$ and density $\rho_{\infty}$ both set to $1$. 
Additionally, 
for unsteady flows, the Strouhal number is given by $St=fD/u_\infty$, 
while the Reynolds number is expressed as $Re=\rho_{\infty}u_{\infty}D/\mu$, with a value of $100$ for a cylinder diameter $D=2$.  
The computational domain is $[15D, 8D]$, 
positioning the cylinder center at $(5D, 4D)$, 
and the final simulation time is $t=200$, 
with the spatial resolutions of $D/dp=10$, $20$, and $40$, 
where the minimal resolution is $dp_{min}=dp/2.0$, 
employed for the convergence study.

\begin{figure}
	\centering
	\includegraphics[width=0.9\textwidth]{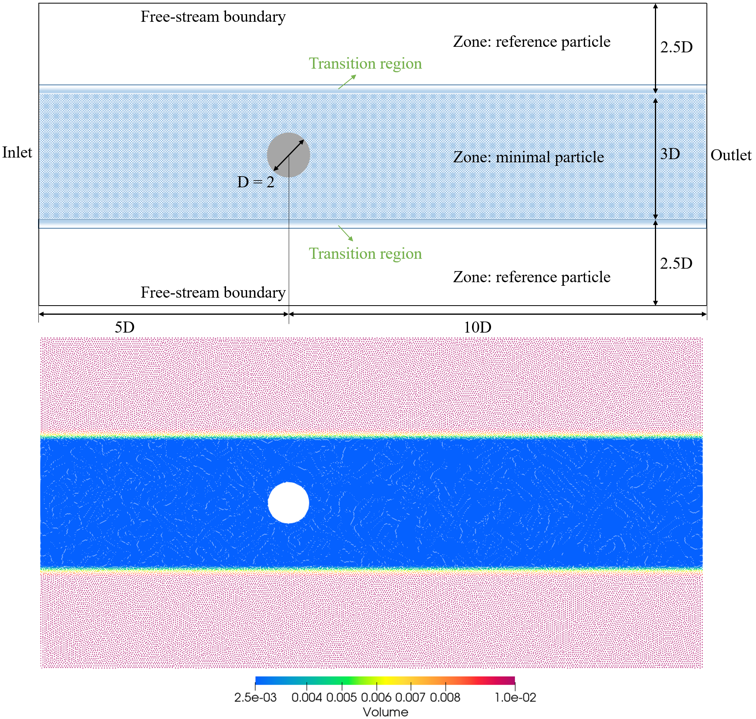}
	\caption{Multi-resolution flow around cylinder ($Re = 100$): 
 Geometry, boundary conditions and the refinement region setup (top panel), as well as particle volume, where the reference particle volume is $0.01$ and the minimal particle volume is $0.0025$, with the resolution of $D/dp=20$ at the initial time (bottom panel).}
	\label{FAC_boundary_setup}
\end{figure}
\begin{figure}
	\centering
	\includegraphics[width=0.9\textwidth]{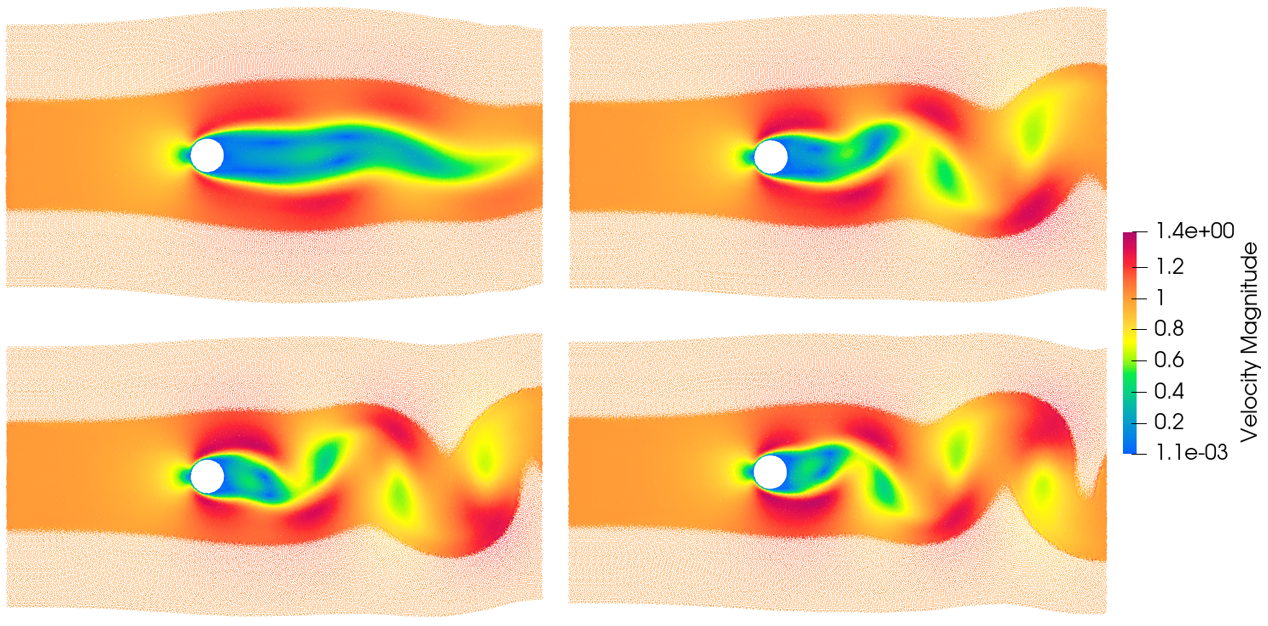}
	\caption{Multi-resolution flow around cylinder ($Re = 100$): 
 velocity contour ranging from $1.1\times 10^{-3}$ to $1.4$ with the spatial resolution of $D/dp=20$ at different time instants.}
	\label{FAC_velocity_different_instants}
\end{figure}
\begin{figure}
	\centering
	\begin{subfigure}[b]{0.49\textwidth}
		\centering
		\includegraphics[trim = 0cm 0cm 0cm 0cm, clip, width=1.0\textwidth]{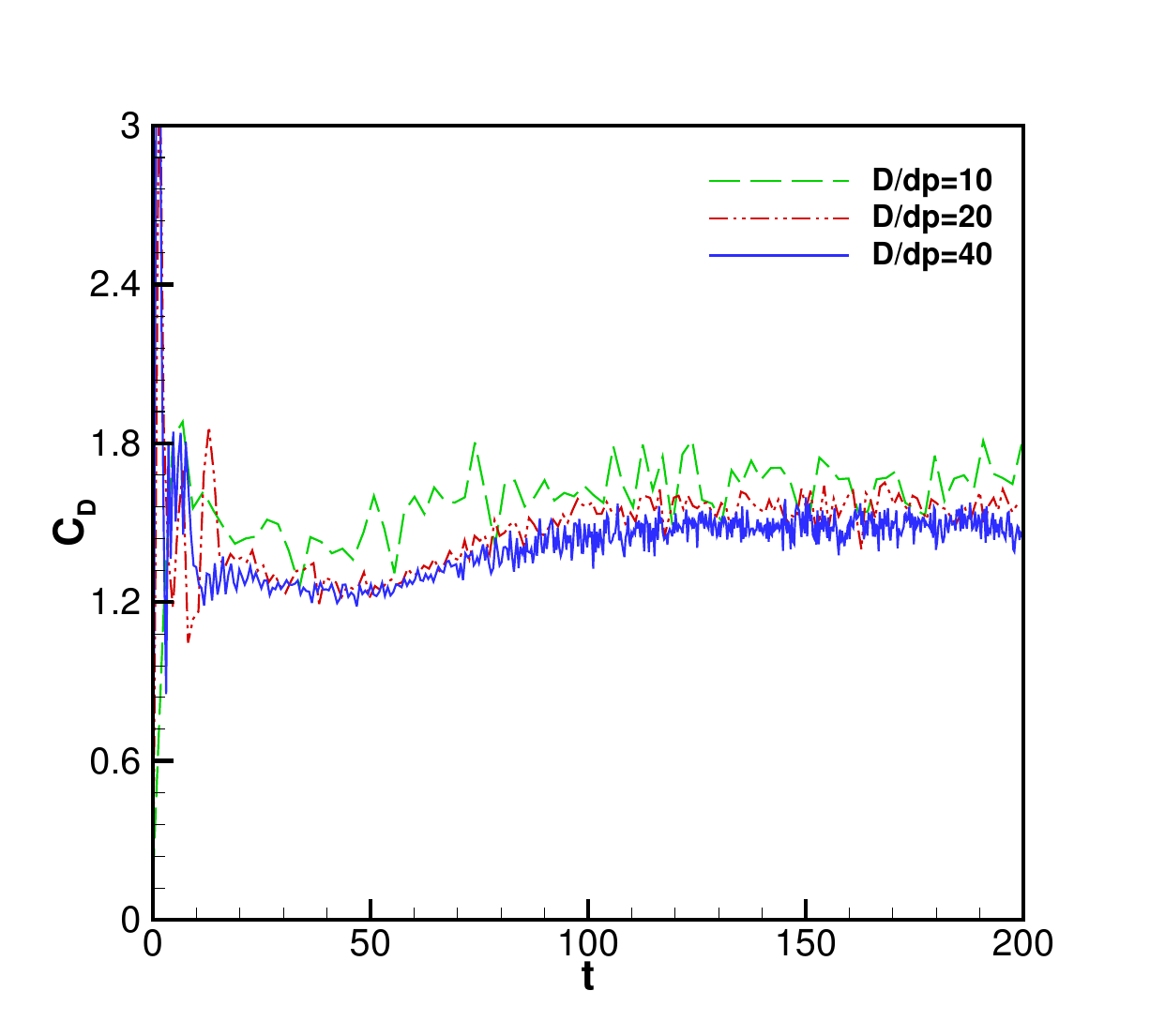}
	\end{subfigure}
	\centering
	\begin{subfigure}[b]{0.49\textwidth}
		\centering
		\includegraphics[trim = 0cm 0cm 0cm 0cm, clip, width=1.0\textwidth]{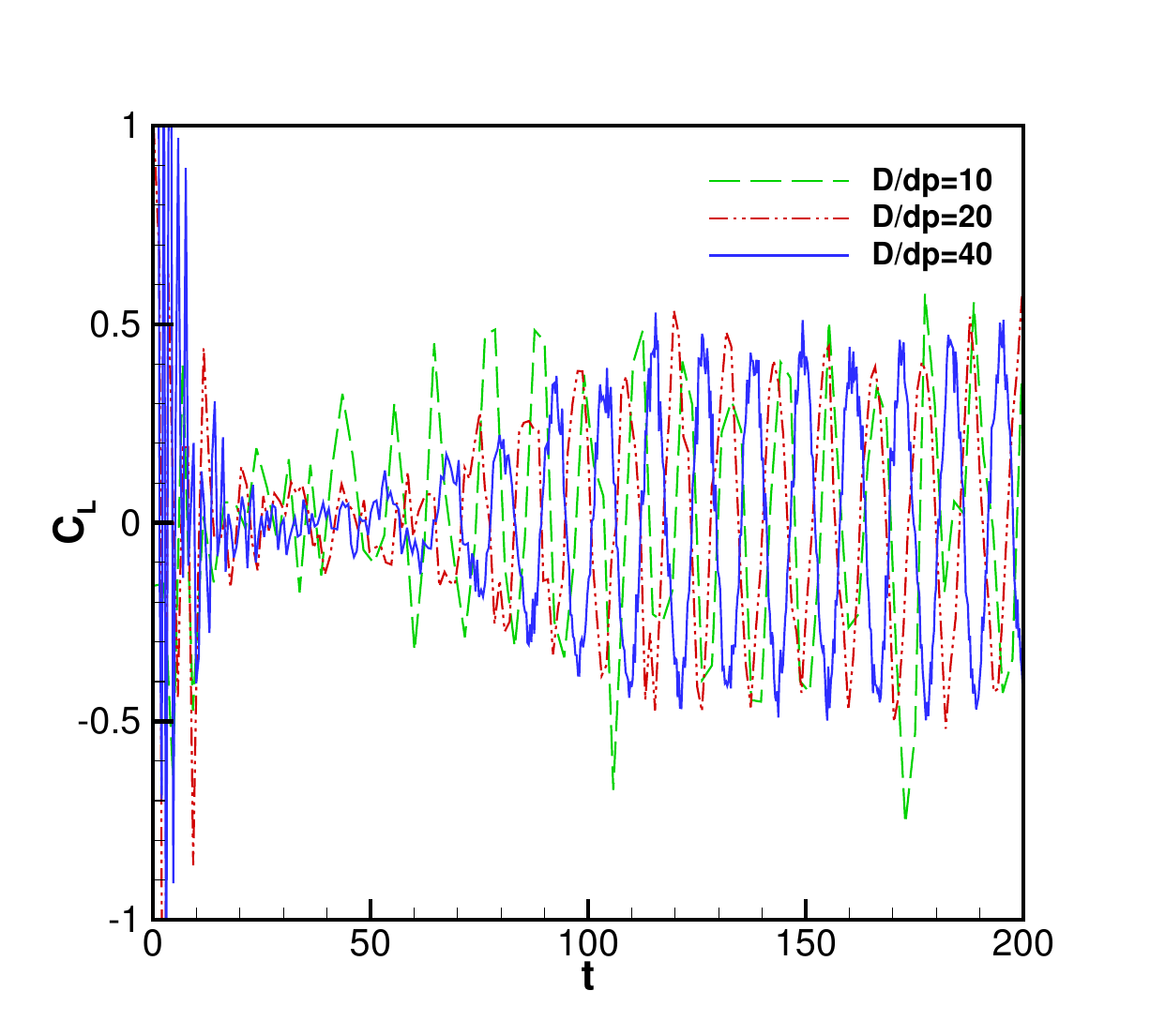}
	\end{subfigure}
	\caption{Multi-resolution flow around cylinder ($Re = 100$): Drag (left panel) and lift (right panel) coefficients with the time using
	 the proposed formulation with $Re=100$.}
	\label{flow around cylinder_coefficients}
\end{figure}
\begin{table}
    \caption{Multi-resolution flow around cylinder ($Re = 100$): Drag and lift coefficients obtained by the proposed formulation with $D/dp=80$ and other experimental and numerical results.}
    \centering
    \begin{tabular}{cccc}
    \hline
    \begin{tabular}[c]{@{}c@{}} Reference\end{tabular} &
    \begin{tabular}[c]{@{}c@{}}$C_{D}$\end{tabular} &
     \begin{tabular}[c]{@{}c@{}}$C_{L}$\end{tabular} &
     \begin{tabular}[c]{@{}c@{}}$S_{t}$\end{tabular} \\ \hline
    White\cite{white2006viscous} & 1.46 & -  & - \\ \hline
    Chiu et al.\cite{chiu2010differentially} & 1.35 $\pm$ 0.012 & $\pm$0.303  & 0.166  \\ \hline
    Le et al.\cite{le2006immersed} & 1.37 $\pm$ 0.009 & $\pm$0.323 & 0.160  \\ \hline
    Brehm et al.\cite{brehm2015locally} & 1.32 $\pm$ 0.010 & $\pm$0.320 & 0.165  \\ \hline
	Russell et al.\cite{russell2003cartesian} & 1.38 $\pm$ 0.007 & $\pm$0.300 & 0.172  \\ \hline
 Zhang et al.\cite{zhang2023lagrangian} & 1.61 $\pm$ 0.005 & $\pm$0.448 & 0.171  \\ \hline
	Present & 1.48 $\pm$ 0.015 & $\pm$0.405 & 0.178  \\ \hline
    \end{tabular}
	\label{Re=100}
\end{table}

Figure \ref{FAC_boundary_setup} portrays the computational geometry, boundary conditions and the refinement region (top panel), 
as well as the particle volume distribution, 
where the reference particle volume is $0.01$ and the minimum particle volume is $0.0025$, 
with the resolution of $D/dp=20$ at the initial time (bottom panel). 
Additionally, 
Figure \ref{FAC_velocity_different_instants} presents the particle velocity contour ranging from $1.1\times 10^{-3}$ to $1.4$ with the resolution of $D/dp=20$. 
This demonstrates a smooth particle distribution and velocity contour. 
To quantitatively assess the accuracy of the proposed formulation, 
the drag coefficient $C_D$ and lift coefficient $C_L$ calculated at a resolution of $D/dp=80$ are compared against reference results in Table \ref{Re=100}. 
These findings indicate that the proposed formulation is effective in variable resolution scenarios.
%
%
\subsection{Three-dimensional lid-driven cavity flow}
\begin{figure}
	\centering
	\begin{subfigure}[b]{0.4\textwidth}
		\centering
		\includegraphics[trim = 0cm 0cm 0cm 0cm, clip, width=1.0\textwidth]{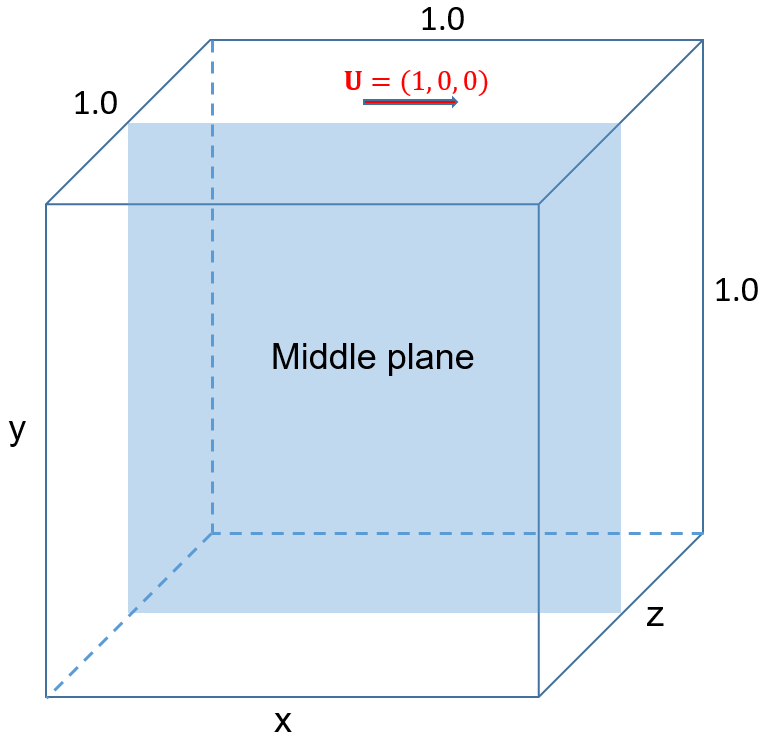}
	\end{subfigure}
	\centering
	\begin{subfigure}[b]{0.49\textwidth}
		\centering
		\includegraphics[trim = 0cm 0cm 0cm 0cm, clip, width=1.0\textwidth]{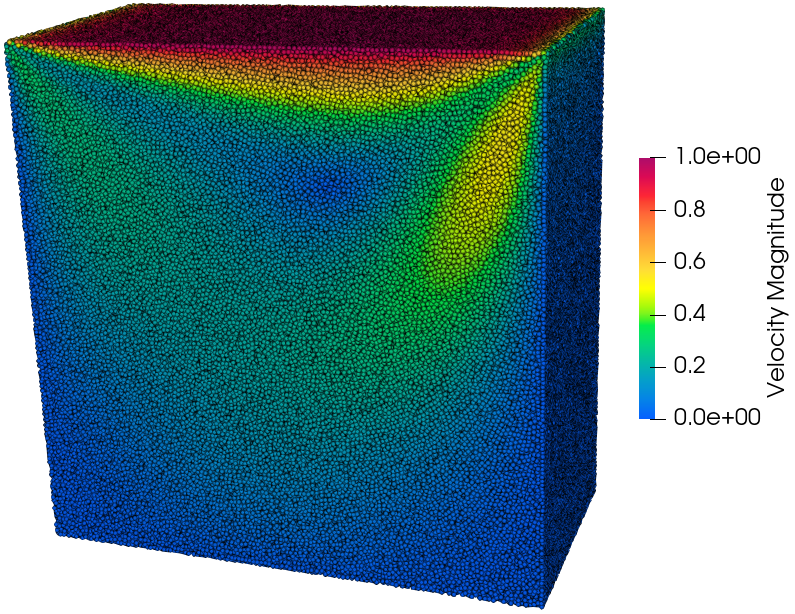}
	\end{subfigure}
	\caption{Three-dimensional lid-driven square cavity flow ($Re=100$): The geometry and boundary conditions setup and the velocity contour of the block after the middle plane ranging from $0$ to $1$ with the resolution of $dp=1/128$.}
	\label{3D Lid-driven cavity flow: Geometry_and_velocity_contour}
\end{figure}

In this section, 
we assess the proposed formulation's effectiveness using a three-dimensional benchmark test, 
specifically a three-dimensional lid-driven cavity flow, 
to validate its performance in 3D scenarios.
As referenced in \cite{ku1987pseudospectral,komaizi2024hybrid}, 
similar to its two-dimensional counterpart, 
the upper boundary is defined as a moving surface with velocity $U=(1.0, 0.0, 0.0)$, 
while all other boundaries are fixed with no-slip conditions, 
as depicted in the left panel of Figure \ref{3D Lid-driven cavity flow: Geometry_and_velocity_contour} (left panel).
In this case, 
the Reynolds number is set as $Re=100$.
\begin{figure}
	\centering
	\begin{subfigure}[b]{0.49\textwidth}
		\centering
		\includegraphics[trim = 0cm 0cm 0cm 0cm, clip, width=1.0\textwidth]{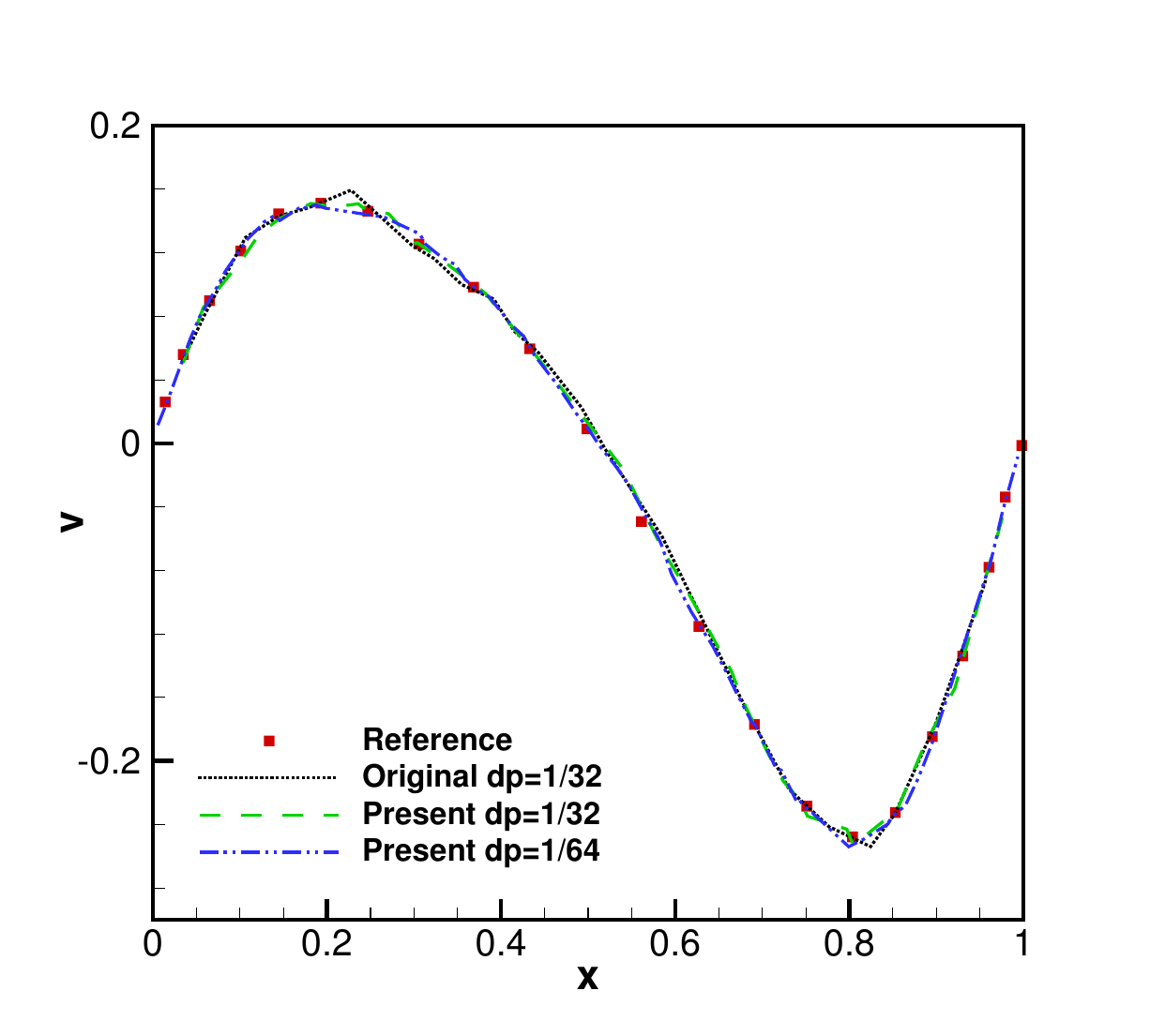}
	\end{subfigure}
	\centering
	\begin{subfigure}[b]{0.49\textwidth}
		\centering
		\includegraphics[trim = 0cm 0cm 0cm 0cm, clip, width=1.0\textwidth]{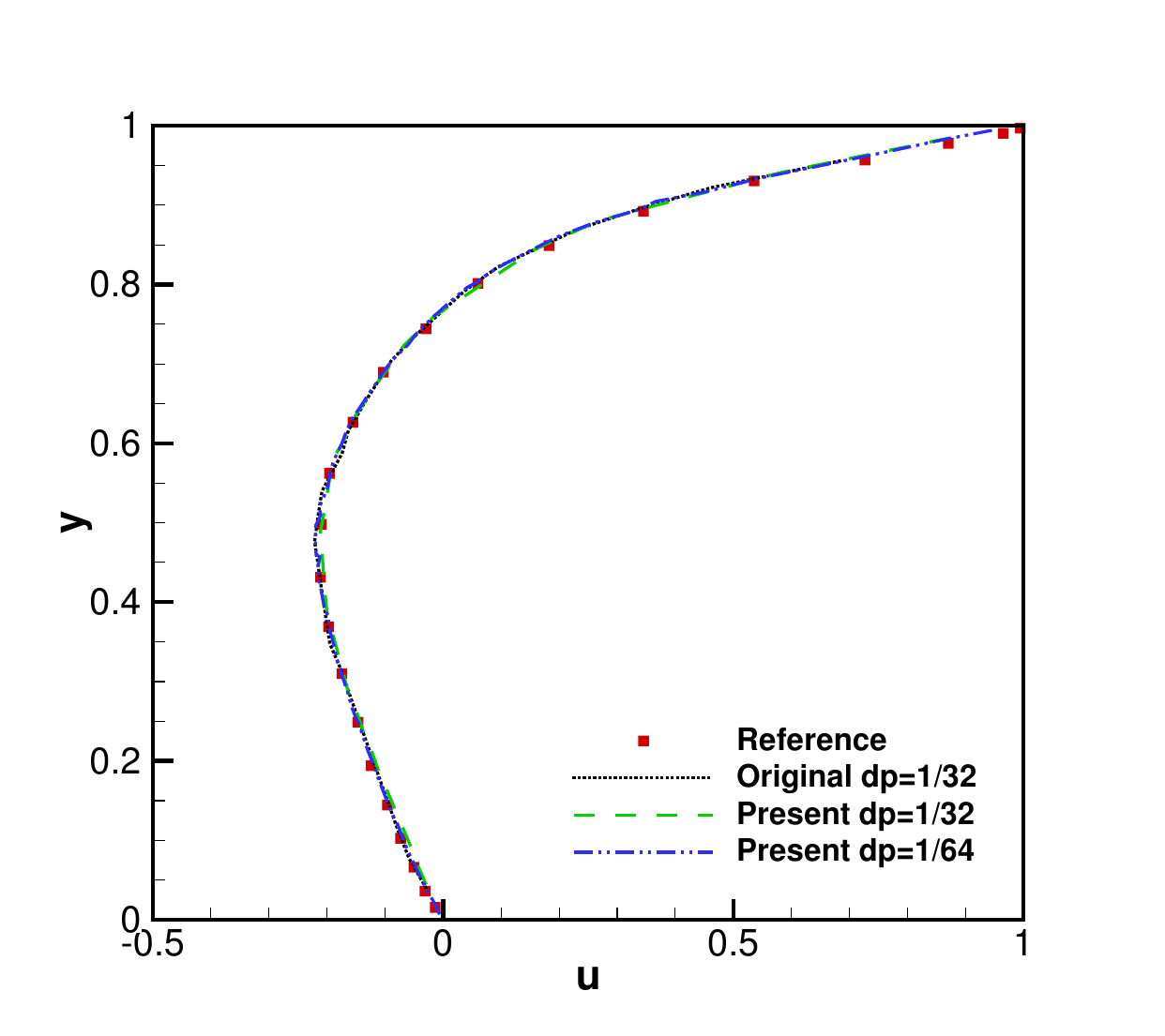}
	\end{subfigure}
	\caption{Three-dimensional lid-driven square cavity flow ($Re=100$): The horizontal (a) and vertical (b) centerlines of the middle plane with the spatial resolutions as $dp=1/32$ and $1/64$, and the comparisons with the reference obtained by Ku et al. \cite{ku1987pseudospectral}.}
	\label{3D_lid-driven-cavity-velocity_profile}
\end{figure}

The right panel of Figure \ref{3D Lid-driven cavity flow: Geometry_and_velocity_contour} presents the velocity contour of the block after the middle plane with the spatial resolution $dp=1/128$, 
demonstrating the ability of the proposed algorithm to achieve smooth results. 
Figure \ref{3D_lid-driven-cavity-velocity_profile} presents the velocity profiles of the horizontal and vertical centerlines of the middle plane with the spatial resolution $dp=1/32$ and $dp=1/64$, 
and compares them with the reference results from \cite{ku1987pseudospectral}. 
The results validate that, in the 3D case, the newly proposed formulation can achieve results comparable to those of the original algorithm, 
and both are in good agreement with the reference result \cite{ku1987pseudospectral}.
%
%
\subsection{Three-dimensional FDA nozzle}
\begin{figure}
	\centering
	\includegraphics[width=0.9\textwidth]{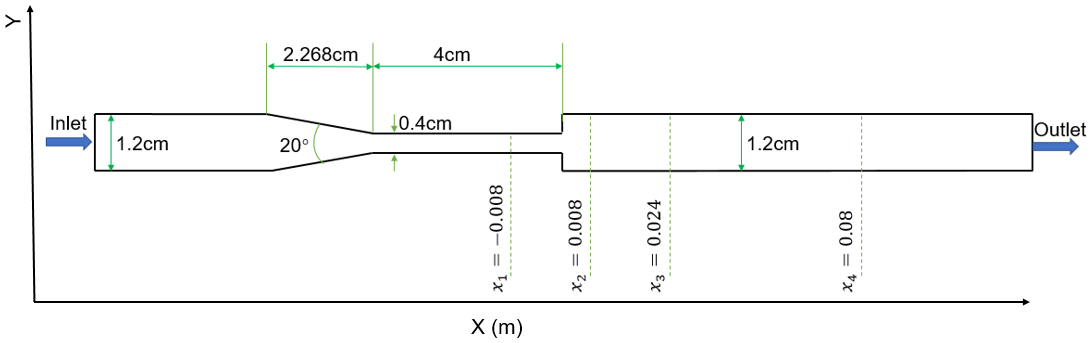}
	\caption{Three-dimensional FDA nozzle: 
		Geometry and boundary conditions.}
	\label{FDA_boundary_setup}
\end{figure}

The section tests a simplified, idealized medical device consisting of a small nozzle to investigate the accuracy of the proposed method. 
Following Refs. \cite{stewart2012assessment, huang2022simulation, hariharan2011experimental}, 
the geometry and boundary conditions are presented in Figure \ref{FDA_boundary_setup}. 
The fluid is specified as Newtonian, with a density of $\rho = 1056 , \text{kg/m}^3$ and dynamic viscosity of $0.0035 , \text{Ns/m}^2$, 
with the Reynolds number in the throat area being $Re = 500$. 
The volume flow rate at the inlet boundary is $Q = 5.2 \times 10^{-6} , \text{m}^3/\text{s}$, 
and the velocity inlet is set as $\overline{U}(r) = 0.5 U_{\text{max}} (1 - r^2/d^2)$, 
where $U_{\text{max}} = U_0 (1.0 - \cos{0.5\pi t})$ if $t \leq t_{\text{ref}}$, 
and $U_{\text{max}} = 2 U_0$ otherwise. 
Here, $U_0$ is the volume flow rate divided by the inlet surface area, and $t_{\text{ref}} = 2.0$. 
In this case, the spatial resolution is set to $dp = 2 \times 10^{-4} , \text{m}$ to discretize the computational domain.

\begin{figure}
	\centering
	\includegraphics[width=0.7\textwidth]{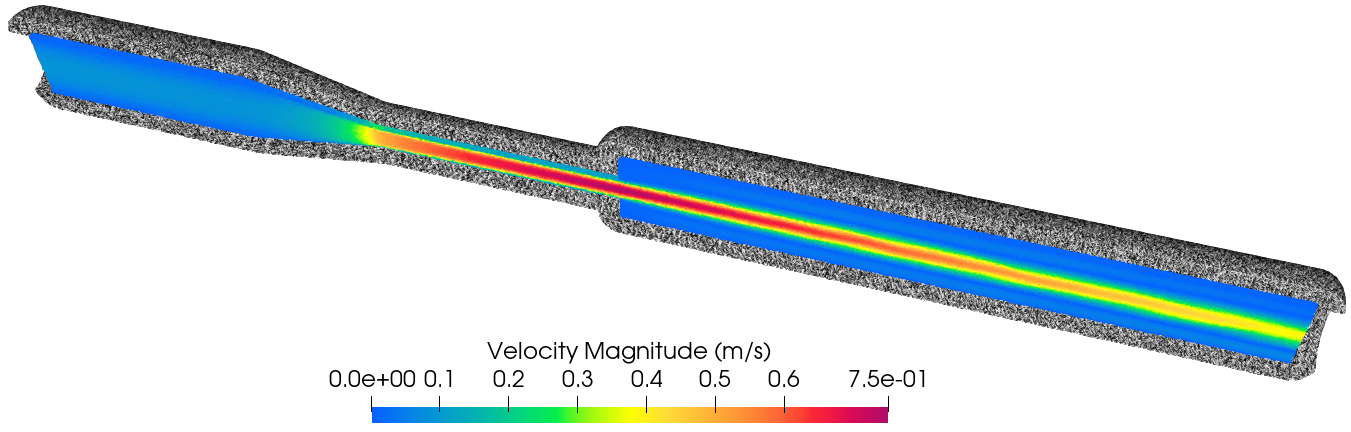}
	\caption{Three-dimensional FDA nozzle (Re=500): 
	Velocity contour using the proposed formulation at time $t=1.6$ s.}
	\label{FDA_velocity_contour}
\end{figure}
\begin{figure}
	\centering
	\includegraphics[width=1.0\textwidth]{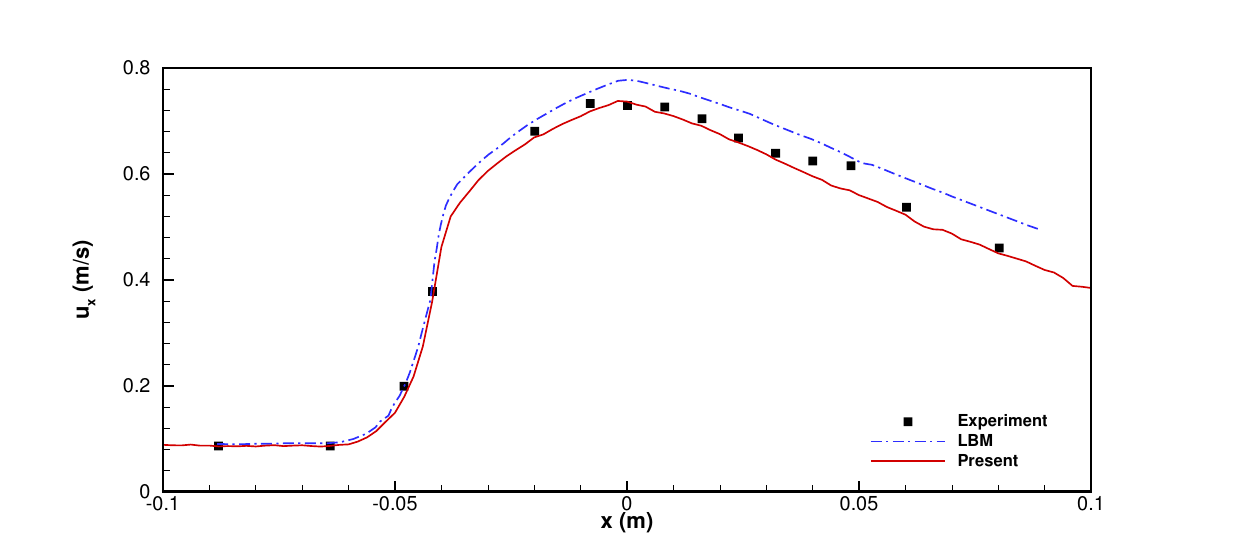}
	\caption{Three-dimensional FDA nozzle (Re=500): 
		Axial velocity using the present algorithm compared with LBM \cite{huang2022simulation} and PIV experiments \cite{hariharan2011experimental} along the center-line of the nozzle.}
	\label{FDA_axis_velocity}
\end{figure}
\begin{figure}
	\centering
	\begin{subfigure}[b]{0.49\textwidth}
		\centering
		\includegraphics[trim = 0cm 0cm 0cm 0cm, clip, width=1.0\textwidth]{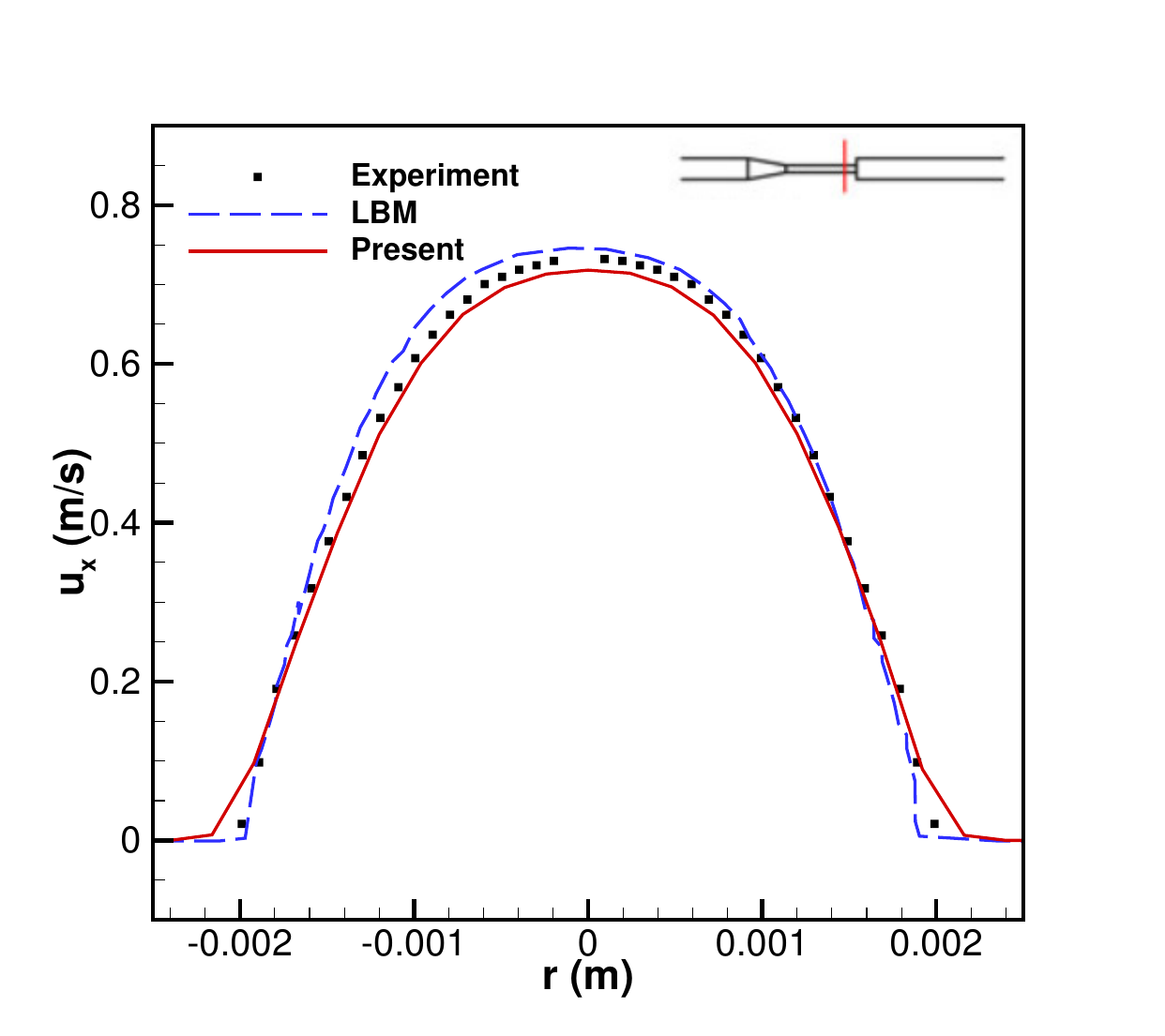}
	\end{subfigure}
	\centering
	\begin{subfigure}[b]{0.49\textwidth}
		\centering
		\includegraphics[trim = 0cm 0cm 0cm 0cm, clip, width=1.0\textwidth]{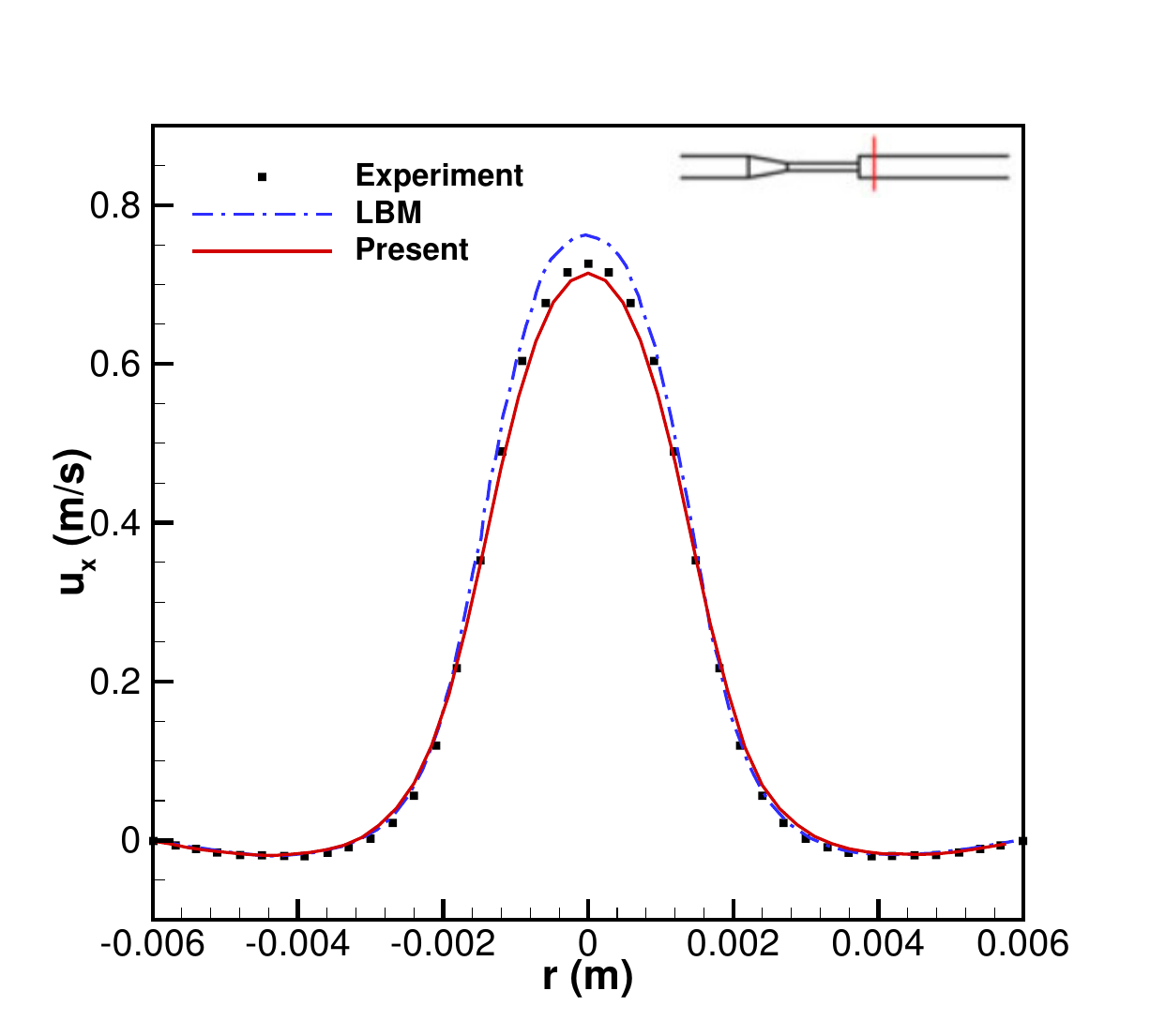}
	\end{subfigure}
\begin{subfigure}[b]{0.49\textwidth}
	\centering
	\includegraphics[trim = 0cm 0cm 0cm 0cm, clip, width=1.0\textwidth]{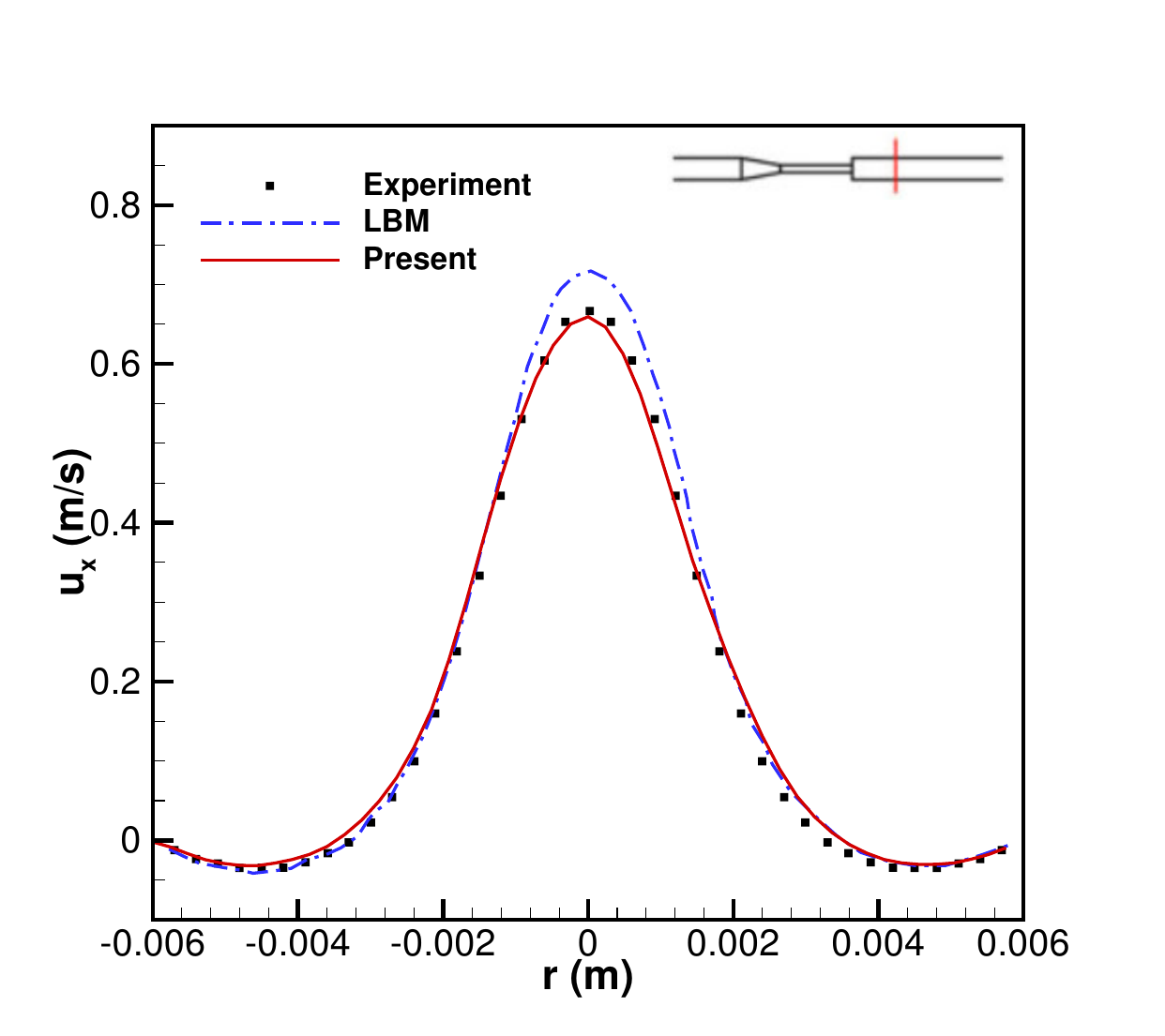}
\end{subfigure}
\begin{subfigure}[b]{0.49\textwidth}
	\centering
	\includegraphics[trim = 0cm 0cm 0cm 0cm, clip, width=1.0\textwidth]{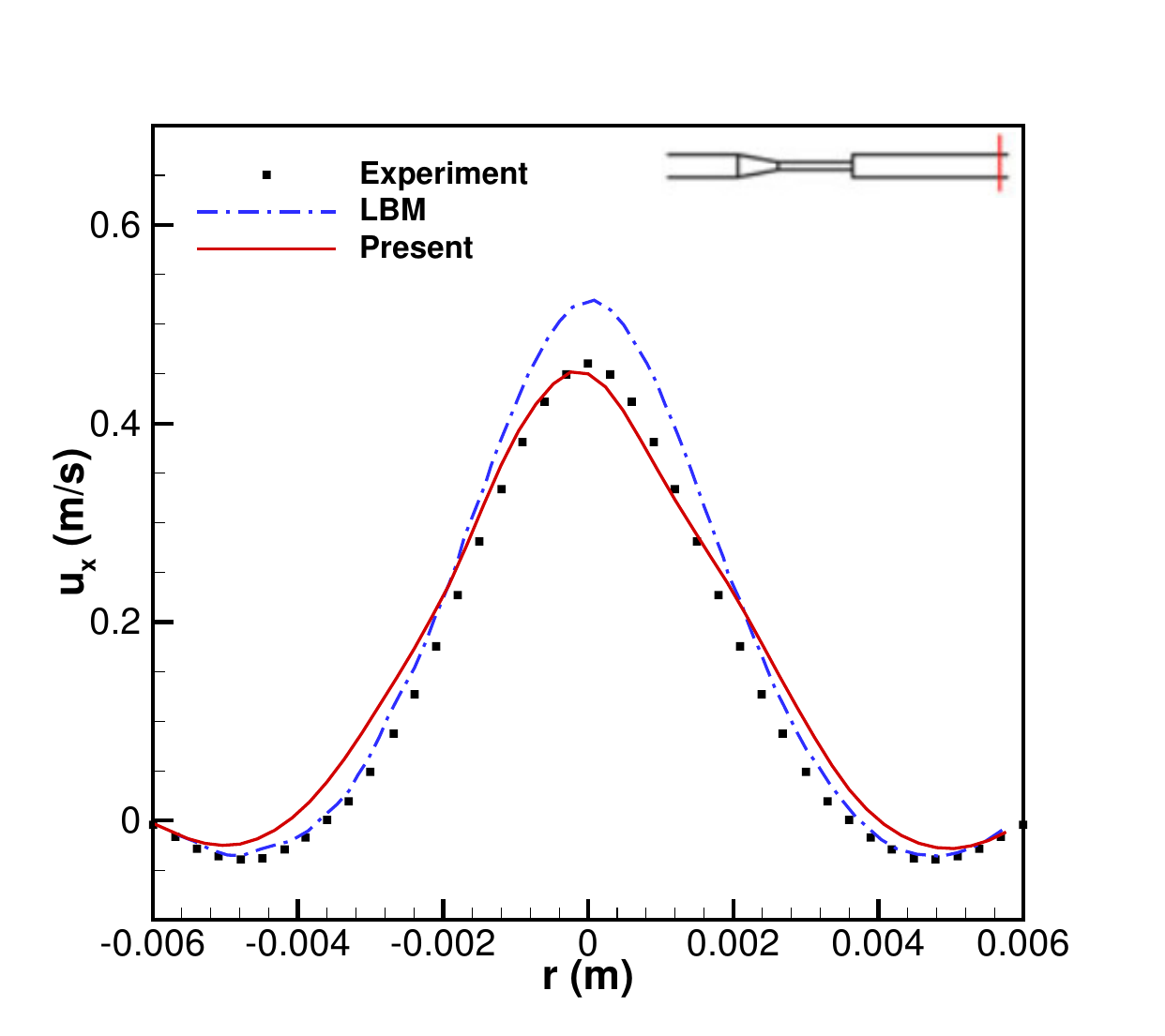}
\end{subfigure}
	\caption{Three-dimensional FDA nozzle (Re=500): 
		Axial velocity profiles from cross-section $x_1$ to $x_4$ using the present algorithm compared with LBM and PIV experiments.}
	\label{FDA_different_cross_section}
\end{figure}

Figure \ref{FDA_velocity_contour} presents the velocity profile using the proposed transport velocity formulation at time $t=1.6s$, 
exhibiting good smoothness. 
Additionally, Figure \ref{FDA_axis_velocity} portrays the axial velocity using the present algorithm compared with the lattice Boltzmann method (LBM) \cite{huang2022simulation}, 
with a resolution of $dp=2.3\times 10^{-4}$, 
and PIV experiments \cite{hariharan2011experimental} along the centerline of the nozzle, 
validating that the result obtained by the present algorithm agrees well with the experimental results. Furthermore, Figure \ref{FDA_different_cross_section} shows the axial velocity profiles from cross-sections $x_1$ to $x_4$ using the present algorithm compared with LBM and PIV experiments. 
The results demonstrate that the proposed formulation achieves consistent results with both the LBM and experiments, proving its validity.
%
%
\section{Summary and conclusion}\label{Summary and conclusions}
In this paper, 
we propose an improved transport velocity correction formulation scaled to the smoothing length instead of background pressure to effectively adapt it to the variable-resolution scheme.
Furthermore, 
an effective limiter is also included for the sceneries with small-velocity to avoid overcorrection.
In addition, 
a series of $2D$ and $3D$ numerical cases have been tested. 
The results have validated that the proposed formulation avoids over-correction for the flow case with small velocity, 
suites the variable-resolution scheme and achieves comparable accuracy using the original formulation.
%
%
\bibliographystyle{elsarticle-num}
\bibliography{ref}

\begin{thebibliography}{10}
\expandafter\ifx\csname url\endcsname\relax
  \def\url#1{\texttt{#1}}\fi
\expandafter\ifx\csname urlprefix\endcsname\relax\def\urlprefix{URL }\fi
\expandafter\ifx\csname href\endcsname\relax
  \def\href#1#2{#2} \def\path#1{#1}\fi

\bibitem{adami2013transport}
S.~Adami, X.~Hu, N.~A. Adams, {A transport-velocity formulation for smoothed
  particle hydrodynamics}, Journal of Computational Physics 241 (2013)
  292--307.

\bibitem{lucy1977numerical}
{A numerical approach to the testing of the fission hypothesis}, author={Lucy,
  Leon B}, The astronomical journal 82 (1977) 1013--1024.

\bibitem{gingold1977smoothed}
R.~A. Gingold, J.~J. Monaghan, {Smoothed particle hydrodynamics: theory and
  application to non-spherical stars}, Monthly notices of the royal
  astronomical society 181~(3) (1977) 375--389.

\bibitem{monaghan2002sph}
J.~J. Monaghan, {SPH compressible turbulence}, Monthly Notices of the Royal
  Astronomical Society 335~(3) (2002) 843--852.

\bibitem{welton1998two}
W.~C. Welton, {Two-dimensional PDF/SPH simulations of compressible turbulent
  flows}, Journal of Computational Physics 139~(2) (1998) 410--443.

\bibitem{monaghan1994simulating}
J.~J. Monaghan, {Simulating free surface flows with SPH}, Journal of
  computational physics 110~(2) (1994) 399--406.

\bibitem{libersky1993high}
L.~D. Libersky, A.~G. Petschek, T.~C. Carney, J.~R. Hipp, F.~A. Allahdadi,
  {High strain Lagrangian hydrodynamics: a three-dimensional SPH code for
  dynamic material response}, Journal of computational physics 109~(1) (1993)
  67--75.

\bibitem{randles1996smoothed}
P.~Randles, L.~D. Libersky, {Smoothed particle hydrodynamics: some recent
  improvements and applications}, Computer methods in applied mechanics and
  engineering 139~(1-4) (1996) 375--408.

\bibitem{longshaw2015automotive}
S.~M. Longshaw, B.~D. Rogers, {Automotive fuel cell sloshing under temporally
  and spatially varying high acceleration using GPU-based Smoothed Particle
  Hydrodynamics (SPH)}, Advances in Engineering Software 83 (2015) 31--44.

\bibitem{wang2023eulerian}
Z.~Wang, C.~Zhang, O.~J. Haidn, X.~Hu, {An Eulerian SPH method with WENO
  reconstruction for compressible and incompressible flows}, Journal of
  Hydrodynamics 35~(2) (2023) 210--221.

\bibitem{khayyer2022systematic}
A.~Khayyer, H.~Gotoh, Y.~Shimizu, {On systematic development of FSI solvers in
  the context of particle methods}, Journal of Hydrodynamics 34~(3) (2022)
  395--407.

\bibitem{monaghan2005smoothed}
J.~J. Monaghan, {Smoothed particle hydrodynamics}, Reports on progress in
  physics 68~(8) (2005) 1703.

\bibitem{schuessler1981comments}
I.~Schuessler, D.~Schmitt, {Comments on smoothed particle hydrodynamics},
  Astronomy and Astrophysics, vol. 97, no. 2, Apr. 1981, p. 373-379. 97 (1981)
  373--379.

\bibitem{johnson1996normalized}
G.~R. Johnson, S.~R. Beissel, {Normalized smoothing functions for SPH impact
  computations}, International Journal for Numerical Methods in Engineering
  39~(16) (1996) 2725--2741.

\bibitem{colagrossi2003numerical}
A.~Colagrossi, M.~Landrini, {Numerical simulation of interfacial flows by
  smoothed particle hydrodynamics}, Journal of computational physics 191~(2)
  (2003) 448--475.

\bibitem{monaghan1989problem}
J.~Monaghan, {On the problem of penetration in particle methods}, Journal of
  Computational physics 82~(1) (1989) 1--15.

\bibitem{xu2009accuracy}
R.~Xu, P.~Stansby, D.~Laurence, {Accuracy and stability in incompressible SPH
  (ISPH) based on the projection method and a new approach}, Journal of
  computational Physics 228~(18) (2009) 6703--6725.

\bibitem{xu2018technique}
X.~Xu, P.~Yu, {A technique to remove the tensile instability in weakly
  compressible SPH}, Computational Mechanics 62~(5) (2018) 963--990.

\bibitem{lind2012incompressible}
S.~J. Lind, R.~Xu, P.~K. Stansby, B.~D. Rogers, {Incompressible smoothed
  particle hydrodynamics for free-surface flows: A generalised diffusion-based
  algorithm for stability and validations for impulsive flows and propagating
  waves}, Journal of Computational Physics 231~(4) (2012) 1499--1523.

\bibitem{antuono2010free}
M.~Antuono, A.~Colagrossi, S.~Marrone, D.~Molteni, {Free-surface flows solved
  by means of SPH schemes with numerical diffusive terms}, Computer Physics
  Communications 181~(3) (2010) 532--549.

\bibitem{sun2017deltaplus}
P.-N. Sun, A.~Colagrossi, S.~Marrone, A.-M. Zhang, {The $\delta$plus-SPH model:
  Simple procedures for a further improvement of the SPH scheme}, Computer
  Methods in Applied Mechanics and Engineering 315 (2017) 25--49.

\bibitem{vacondio2013variable}
R.~Vacondio, B.~D. Rogers, P.~K. Stansby, P.~Mignosa, J.~Feldman, {Variable
  resolution for SPH: a dynamic particle coalescing and splitting scheme},
  Computer Methods in Applied Mechanics and Engineering 256 (2013) 132--148.

\bibitem{lyu2022further}
H.-G. Lyu, P.-N. Sun, {Further enhancement of the particle shifting technique:
  Towards better volume conservation and particle distribution in SPH
  simulations of violent free-surface flows}, Applied Mathematical Modelling
  101 (2022) 214--238.

\bibitem{zhang2017weakly}
C.~Zhang, X.~Hu, N.~A. Adams, {A weakly compressible SPH method based on a
  low-dissipation Riemann solver}, Journal of Computational Physics 335 (2017)
  605--620.

\bibitem{litvinov2015towards}
S.~Litvinov, X.~Hu, N.~A. Adams, {Towards consistence and convergence of
  conservative SPH approximations}, Journal of Computational Physics 301 (2015)
  394--401.

\bibitem{zhang2017generalized}
C.~Zhang, X.~Y. Hu, N.~A. Adams, {A generalized transport-velocity formulation
  for smoothed particle hydrodynamics}, Journal of Computational Physics 337
  (2017) 216--232.

\bibitem{zhang2020dual}
C.~Zhang, M.~Rezavand, X.~Hu, {Dual-criteria time stepping for weakly
  compressible smoothed particle hydrodynamics}, Journal of Computational
  Physics 404 (2020) 109135.

\bibitem{zhang2021sphinxsys}
C.~Zhang, M.~Rezavand, Y.~Zhu, Y.~Yu, D.~Wu, W.~Zhang, J.~Wang, X.~Hu,
  {SPHinXsys: An open-source multi-physics and multi-resolution library based
  on smoothed particle hydrodynamics}, Computer Physics Communications 267
  (2021) 108066.

\bibitem{toro2013riemann}
E.~F. Toro, {Riemann solvers and numerical methods for fluid dynamics: a
  practical introduction}, Springer Science \& Business Media, 2013.

\bibitem{hu2006multi}
X.~Y. Hu, N.~A. Adams, {A multi-phase SPH method for macroscopic and mesoscopic
  flows}, Journal of Computational Physics 213~(2) (2006) 844--861.

\bibitem{monaghan1992smoothed}
J.~J. Monaghan, {Smoothed particle hydrodynamics}, Annual review of astronomy
  and astrophysics 30~(1) (1992) 543--574.

\bibitem{zhang2023lagrangian}
S.~Zhang, W.~Zhang, C.~Zhang, X.~Hu, {A Lagrangian free-stream boundary
  condition for weakly compressible smoothed particle hydrodynamics}, Journal
  of Computational Physics 490 (2023) 112303.

\bibitem{wendland1995piecewise}
H.~Wendland, {Piecewise polynomial, positive definite and compactly supported
  radial functions of minimal degree}, Advances in computational Mathematics 4
  (1995) 389--396.

\bibitem{taylor1937mechanism}
G.~I. Taylor, A.~E. Green, {Mechanism of the production of small eddies from
  large ones}, Proceedings of the Royal Society of London. Series
  A-Mathematical and Physical Sciences 158~(895) (1937) 499--521.

\bibitem{ghia1982high}
U.~Ghia, K.~N. Ghia, C.~Shin, {High-Re solutions for incompressible flow using
  the Navier-Stokes equations and a multigrid method}, Journal of computational
  physics 48~(3) (1982) 387--411.

\bibitem{turek2006proposal}
S.~Turek, J.~Hron, {Proposal for numerical benchmarking of fluid-structure
  interaction between an elastic object and laminar incompressible flow},
  Springer, 2006.

\bibitem{han2018sph}
L.~Han, X.~Hu, {SPH modeling of fluid-structure interaction}, Journal of
  Hydrodynamics 30 (2018) 62--69.

\bibitem{zhang2021multi}
C.~Zhang, M.~Rezavand, X.~Hu, {A multi-resolution SPH method for
  fluid-structure interactions}, Journal of Computational Physics 429 (2021)
  110028.

\bibitem{bhardwaj2012benchmarking}
R.~Bhardwaj, R.~Mittal, {Benchmarking a coupled
  immersed-boundary-finite-element solver for large-scale flow-induced
  deformation}, AIAA journal 50~(7) (2012) 1638--1642.

\bibitem{tian2014fluid}
F.-B. Tian, H.~Dai, H.~Luo, J.~F. Doyle, B.~Rousseau, {Fluid--structure
  interaction involving large deformations: 3D simulations and applications to
  biological systems}, Journal of computational physics 258 (2014) 451--469.

\bibitem{white2006viscous}
F.~M. White, J.~Majdalani, {Viscous fluid flow}, Vol.~3, McGraw-Hill New York,
  2006.

\bibitem{chiu2010differentially}
P.-H. Chiu, R.-K. Lin, T.~W. Sheu, {A differentially interpolated direct
  forcing immersed boundary method for predicting incompressible Navier--Stokes
  equations in time-varying complex geometries}, Journal of Computational
  Physics 229~(12) (2010) 4476--4500.

\bibitem{le2006immersed}
D.-V. Le, B.~C. Khoo, J.~Peraire, {An immersed interface method for viscous
  incompressible flows involving rigid and flexible boundaries}, Journal of
  Computational Physics 220~(1) (2006) 109--138.

\bibitem{brehm2015locally}
C.~Brehm, C.~Hader, H.~F. Fasel, {A locally stabilized immersed boundary method
  for the compressible Navier--Stokes equations}, Journal of Computational
  Physics 295 (2015) 475--504.

\bibitem{russell2003cartesian}
D.~Russell, Z.~J. Wang, {A Cartesian grid method for modeling multiple moving
  objects in 2D incompressible viscous flow}, Journal of Computational Physics
  191~(1) (2003) 177--205.

\bibitem{ku1987pseudospectral}
H.~C. Ku, R.~S. Hirsh, T.~D. Taylor, {A pseudospectral method for solution of
  the three-dimensional incompressible Navier-Stokes equations}, Journal of
  Computational Physics 70~(2) (1987) 439--462.

\bibitem{komaizi2024hybrid}
D.~Komaizi, A.~Niknam, {Hybrid finite-volume--finite-element scheme for 3D
  simulation of thermal plasma arc configuration}, AIP Advances 14~(1) (2024).

\bibitem{stewart2012assessment}
S.~F. Stewart, E.~G. Paterson, G.~W. Burgreen, P.~Hariharan, M.~Giarra,
  V.~Reddy, S.~W. Day, K.~B. Manning, S.~Deutsch, M.~R. Berman, et~al.,
  {Assessment of CFD performance in simulations of an idealized medical device:
  results of FDA’s first computational interlaboratory study}, Cardiovascular
  Engineering and Technology 3 (2012) 139--160.

\bibitem{huang2022simulation}
{Simulation of the FDA nozzle benchmark: A lattice Boltzmann study},
  author={Huang, Feng and No{\"e}l, Romain and Berg, Philipp and Hosseini,
  Seyed Ali}, Computer Methods and Programs in Biomedicine 221 (2022) 106863.

\bibitem{hariharan2011experimental}
P.~Hariharan, M.~Giarra, V.~Reddy, S.~Day, K.~Manning, S.~Deutsch, S.~Stewart,
  M.~Myers, M.~Berman, G.~Burgreen, et~al., Experimental particle image
  velocimetry protocol and results database for validating computational fluid
  dynamic simulations of the fda benchmark nozzle model, Journal of
  Biomechanical Engineering 133 (2011) 041002.

\end{thebibliography}
\end{document}